# Energy *Realpolitik*: Towards a Sustainable Energy Strategy

W. Udo Schröder

*Departments of Chemistry and Physics*
*University of Rochester, Rochester, N.Y. 14627*
[Schroeder@chem.rochester.edu](mailto:Schroeder@chem.rochester.edu)

## Abstract

*A long-term strategy based on existing technological, ecological, economical, and geopolitical realities is urgently needed to develop a sustainable energy economy, which should be designed with adaptability to unpredicted changes in any of these aspects. While only a highly diverse energy portfolio and conservation can ultimately guarantee optimum sustainability, based on a comparison of current primary energy generation methods, it is argued that future energy strategy has to rely heavily on expanded coal and nuclear energy sectors. A comparison of relative potentials, merits and risks associated with fossil-fuel, renewable, and nuclear technologies suggests that the balance of technologies should be shifted in favor of new-generation, safe nuclear methods to produce electricity, while clean-coal plants should be assigned to transportation fuel. Novel nuclear technologies exploit fission of uranium and thorium as primary energy sources with fast-spectrum and transmutation (burner) reactors. A closed fuel cycle and waste transmutation resolve the strategic issues associated with nuclear power. Innovative reactor designs utilize spallation of heavy metals in subcritical accelerator driven systems or molten-salt reactors. Importation and reconstruction of technical expertise already lost and aversion of further erosion are preconditions to any successful energy strategy. Research opportunities to perfect innovative nuclear, coal, and renewable energy technologies should be pursued.*

## Introduction

As the Earth's population and its aspirations for a healthier and more prosperous standard of living grow, requirements of continuous, clean and environmentally sustainable primary energy production increase dramatically. Typical estimates envision the global demand to increase from ~12 TW today, 85% of which is fossil-fueled, to more than ~45TW by mid-century. Earth's dwindling energy resources and insufficient basic energy infrastructure give ample reason to doubt that such enormous demands could actually be satisfied. Increasing international competition for access to energy resources adds reason to be concerned about the degree to which goals of national energy independence and security can be approached. Proposed energy strategies lead to the conclusion that the production ideals stated above are virtually unattainable. In particular, the U.S. fraction (25%) of total world energy consumption has to decline significantly. Therefore, parallel strategic action is needed to achieve significant reduction in energy demand, a much improved energy distribution infrastructure (grid), and new energy R&D. Within this framework, a sustainable strategy for generating primary energy is crucial, one which

- has good prospects to satisfy a variable mix of energy demands by flexible response;
- enhances energy independence;
- takes into account existing infrastructure and its evolutionary inertia;
- is largely independent of future technological breakthroughs; and
- is acceptable to the public.

It follows from these principles that in the palette of energy technologies those primary energy resources should be developed with highest priority that

1. are domestically abundant and economically accessible with current technologies,
2. tie in with existing distribution networks,
3. satisfy a steady base load demand,
4. are safe to produce and use,
5. are adaptable to a variety of fuel demands,
6. minimize the release of pollutants.

The following outlines and puts in perspective aspects of a long-term strategy for securing a reliable primary energy economy, realizing modern nuclear and "clean-coal" technologies, which appear to be the only serious contenders for the task.

**Task definition and background**

The basic energy supply task faced by industrialized and developing nations alike is complicated by simultaneous calls for an overall limitation, and even a reduction, in the associated output of "greenhouse gases" (GHG). Anthropogenic greenhouse gases, most importantly carbon dioxide ($CO_2$), methane ($CH_4$), nitric gases ($NO_x$, x = 1, 2), and chlorofluorocarbons ($CFC$), but also water vapor ($H_2O$), are suspected to influence the global climate in an undesirable way, thereby harming the biosphere and threatening specifically human civilizations. Additional complexity arises from the fact that mobile and stationary energy consuming sectors of most economies require different approaches in terms of primary and secondary energy carriers and distribution networks. While stationary energy demands can, in principle, be satisfied by electricity supplied by the grid, the prevalent transportation fleet largely requires fuel in convenient onboard packages, presently realized by liquid fuels with high energy (caloric) density. To maintain such complexity under future variable external conditions requires a highly flexible domestic energy infrastructure based on a diverse set of primary energy sources. In practice, this goal can be realized only in stages.

The public has been sensitized to aspects of energy scarcity and environmental problems by rising oil and gas prices, widely publicized scenarios of impending climatic changes, and evidently insufficient urban air quality. But it has largely been left unaware of the technological boundary conditions imposed on a realistic, flexible and sustainable energy strategy and its coupling to the existing economic infrastructure.

Well publicized issues associated with fossil or nuclear fuel based energy alternatives include the limited availability of the primary fuels, the environmental impact and risks in mining and transport, and other hazards associated with plant security and operations. Major objections to these technologies relate to the disposition of toxic and GHG waste produced by fossil fuel burners and of the long-lived radio-toxins generated by nuclear power plants. The media have played a pernicious role in social amplification of some of these concerns, specifically concerning radiation, supporting irrational attitudes. The opposite is required, a sober look at the available alternatives.

A number of popular science book publications (e.g., Gore, 2006, Morris, 2006) as well as public initiatives and promotions leave the impression that the looming energy/climate crisis is solvable by harnessing increasing amounts of renewable hydroelectric, wind, solar, geothermal and biomass energy sources, which are said to be available, virtually free and ecologically benign. Supposedly, renewable energy technologies will be able to displace those based on fossil and nuclear fuels, which are deemed ecologically disastrous and therefore to be phased out.

Discussion in the present article supports a different and more realistic view promoted to some extent also by other studies of possible and desirable energy futures (Hoffert, 2002, Deutch, 2003, Romm, 2005). The aim of such studies is to assess and inform government energy policies in the U.S. and other industrialized countries around the world. There are several fundamental and practical reasons for insisting on energy *Realpolitik*, i.e., a responsible energy policy oriented at present reality, a goal of maximum flexibility and realistic short term extrapolations in feasibility.

On the fundamental level, it has been known since the latter half of the 20[th] century that complex or "emerging," strongly coupled, multi-dimensional systems show a non-linear overall response to changes in certain variables, notably the so-called order parameters. Simple mathematical models illustrate (e.g., Mainzer, 1997) how predictable (orderly) behavior observed for certain domains of order parameter(s) can suddenly switch to unpredictable (chaotic) dynamics, once parameters exceed certain domain boundaries.

Even a minimalistic view has to admit that the global climate and national (or global) economics (Krugman, 1996, Schweitzer, 2002) belong to the category of complex systems capable of orderly and chaotic dynamics. Neither is understood in enough detail to allow accurate predictions to be made for their dynamical evolution beyond the short term of, perhaps, a few years or a decade. On the other hand, plausible arguments have been made (IPCC, 2007) that anthropogenic actions can lead to a coupling of global climate and economy creating a supercomplex coupled system. Coupling factors include large-scale deforestation and atmospheric GHG emissions. In the interest of a better predictable (stable) global future, it then makes sense to strive to reduce anthropogenic coupling of climate and economic systems, i.e., to reduce the "carbon footprint" of the economy.

On the practical level, the success of past future forecasts does not instill great confidence in present attempts at this art (Smil, 2005). Based on these arguments and taking into account the energy infrastructure already existing in industrial nations, a responsible energy strategy for the medium term of the next few decades would emphasize a mix of existing primary energy technologies, but further develop and refine them. In this mix, efficient technologies based on domestically abundant fossil and nuclear fuels would necessarily dominate. Although the long-range potential of fusion energy technology is important for future generations, few energy experts expect that a practical fusion energy power plant can be commissioned long before the end of this century (IFEO, 2004). In the interest of developing long-term diversity, research and development of renewable fuel technologies derived from hydrodynamic (water) flow, solar radiation, wind, geothermal, or biomass, should be encouraged but not relied upon before proven maturity.

It is therefore safe to conclude that the lion share of any future growth in energy use will likely be covered by fossil fuel power plants burning coal and, to a decreasing extent, oil and natural gas (Deutch, 2003, Romm, 2005). Conventional (fission) nuclear power contributing now ~20% worldwide to electricity generation is seen (Deutch et al, 2003) to increase its share to at least ~30% within the next 20 years, although a share of 70-80% such as in France could be attained on a slightly longer time scale.

The energy intensive transportation sector shows trends of increasing electric/liquid fuel hybridization and has generated renewed interest in synthetic fuels. In principle, the expected increased demand for synthetic fuel could be satisfied employing available, albeit expensive and still relatively inefficient, industrial processes of coal gasification, pyrolysis, direct and indirect coal liquefaction (Probstein, 2006, Olah, 2006). Resulting shortages in coal based electricity would have to be mitigated by increased electricity from nuclear power. This likely scenario requires timely preparation, massive investments, as well as further R&D. Should renewable energy technologies mature and overcome their present technological barriers, they could see an overall contribution to the total energy economy higher than the few per-cent now predicted (Romm, 2003, Eerkens, 2006).

In advocating the development of a realistic and sustainable energy strategy for the industrialized world and emerging economies, the present discussion will specifically elaborate arguments for a priority deployment of nuclear power technologies. The merits of nuclear energy have been greatly and consistently underappreciated, while its hazards have been overstated. Although the main alternative energy sources for the short and intermediate term remain oil, natural gas, and nuclear energy, the following includes also a brief account of environmental and technical issues related to renewable energy sources. It is merely intended to place energy alternatives in perspective, a task benefiting from several useful critical reviews (Hoffert, 2002, Deutch, 2003, Ewing, 2004). A thorough critical assessment of the merits of each of these latter renewable energy technologies is duly left in the responsibility of corresponding experts in the respective energy subfields.

**Potential and Risk Assessment: Fossil Fuels**

The conventional fossil fuels comprise natural gas, oil in different consistencies, and different types (ranks) of coal. The U.S. presently consumes daily 3.6 million short tons of coal, 22 million barrels of oil and 63 billion cf of natural gas (EIA, 2006). Coal ranks from low-quality lignite to carbon-rich and dense anthracite. In addition, with the decline of world resources in liquid crude oil, more unconventional sources like shale oil and tar sands are exploited. In the longer term, abundant methane hydrates may also assume an important role in the global energy economy.

Fossil fuels consist basically of complex hydrocarbon chain molecules ($C_nH_m$). They provide energy through combustion, a process in which chemical energies $\Delta H$ of molecular bonding are set free by molecular decomposition and atomic rearrangement. Rearrangement reactions of the type

$$C_nH_m + (n+m/4)O_2 \rightarrow nCO_2 + (m/2)H_2O + \Delta H \quad (1)$$

proceed under heat and in the presence of oxygen ($O_2$), where $n$ and $m$ are stoichiometric coefficients. Besides their function as primary energy sources, such complex organic compounds serve as feedstock for the petrochemical industry, used to make anything from fertilizer to plastics and pharmaceuticals. In addition, with declining crude oil resources, petroleum for the transportation sector may have to be replaced partially by synthetic fuels converted from coal, low-grade oil shale or tars (Probstein, 2006).

An increased long-term use of fossil fuels, in particular coal, faces opposition because of several *Strategic Issues*:

A. Dangers associated with production;
B. Environmental hazards and damage in mining/production;
C. Emission of airborne pollutants (particles, metals, GHG), smog;
D. Uncertainty about carbon capture and permanent sequestration;
E. Limitation of resources, economy.

**A.  Coal**

The most valuable, high-energy density coals have been formed over time spans of several hundred million years (Carboniferous Period). In the U.S., the great Appalachian coal bed is a result of long-term accumulation, compression and heating of organic waste and its transformation into coal of high density and high carbon content. These resources are now largely exhausted in the U.S., as well as in the rest of the industrialized world. Their production has also declined for environmental reasons. On the other hand, coal found, e.g., in the Illinois Basin, in Wyoming and Montana (Powder River Basin) is of much younger age and lesser energy content. The corresponding heat contents range from 14,000 Btu/lb for anthracite down to app. 5,000 Btu/lb for lignite (brown coal).

As shown in Fig. 1 (EPA, 2007), the total consumption of coal in the U.S. has been on the increase since the 1970s. For orientation, in 2005 the U.S. consumed 1.3 Billion short tons of coal (EIA, 2005) corresponding to 23 quadrillion Btu ("quad"), mostly (97%) to generate electricity (~300 GW$_e$) with more than 1,500 coal-fired power plants. It has been estimated that domestic resources can satisfy U.S. demand for periods ranging from 80 to 250 years (Goodell, 2006, EIA, 2007) and beyond, depending on the assumed future economic growth, the associated extent and diversity of coal uses, as well as the tolerated price for production (mining and transport) and retail.

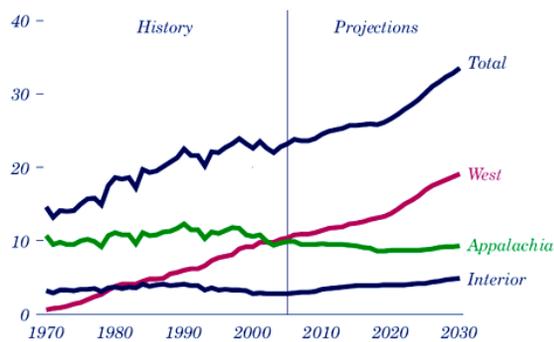

**Figure 1: U.S. coal production (quad. Btu) by geographic region.**

Table 1 gives a summary of (rough) estimates (EIA 2006) of world coal resources deemed recoverable with modern technologies. In addition to the U.S., countries such as India, China, Australia, Russia, and South Africa possess sizeable coal resources, although these coals are often of low quality. In particular the emerging economies of China and India are predicted to fuel their industries during the early parts of this century mainly by burning coal. China is said to construct every week coal-fired power plants in the equivalent of 1 GW.

In view of the facts that a) there are substantial world and domestic coal reserves, b) efficient coal infrastructure is available in the U.S (as in many other countries), and c) a number of new uses are envisioned for coal, the future increase in U.S. coal production anticipated in the energy outlook of Fig. 1 appears quite realistic. However, rapidly rising oil prices and diversification of coal use will require an even steeper rise in coal production.

As will be illustrated below, renewable energy alternatives that would be potent and abundant enough to displace coal from a future dominating position as primary domestic energy source have not (yet) been found. However, increased utilization of nuclear power can mitigate the anticipated surge in coal burning and conversion of coal into synthetic fuels.

Turning now to the risks (Strategic Issues) associated with the production and use of coal as primary energy source, it is widely recog-nized that traditional coal mining belongs to the most dangerous human occupations that still claim many thousands of lives every year, often through underground methane explosions, water flooding or collapse of caves. For example, the U.S Mining Safety and Health Administration (MSHA, 2006) reports in one year 42 fatalities in U.S. coal mines alone, rivaling yearly the singular (1987) Chernobyl nuclear accident. In addition, there are additional health risks in coal mining, e.g., causing respiratory diseases such as silicosis ("black lung").

**Table 1: World recoverable coal reserves (after EIA, 2006)**

| Region/Country | Anthracite Bituminous | Lignite Subbituminous | Total |
|---|---|---|---|
| United States | 123,834 | 143,478 | 267,312 |
| North America | 128,608 | 147,491 | 276,100 |
| Central & South America | 8,489 | 13,439 | 21,928 |
| Western Europe | 1,571 | 34,918 | 36,489 |
| Europe | 19,558 | 46,203 | 65,762 |
| Eurasia | 104,183 | 146,322 | 250,505 |
| Middle East | 462 | 0 | 462 |
| Africa | 55,294 | 192 | 55,486 |
| Asia & Oceania | 212,265 | 114,999 | 327,264 |
| World Total | 528,860 | 468,646 | 997,506 |

In the U.S., most of the underground mining has given way to surface "strip" or open-pit mining of the younger coals of the West. The technique, which has also been used in Eastern Europe and Asia, has enormous environmental impact, on the topography, the fauna and the flora. In such open-cast mining, massive amounts of surface soil and dirt ("overburden") are removed by giant machines or by dynamite blasts, leveling mountain tops and filling in valleys and river beds. In the course of this type of mining, animal habitat is destroyed along with the sustaining vegetation. Polluted acidic runoff from the mine sites, unimpeded by trees and other vegetation, poisons local rivers, creeks, and the main supplies of drinking water with (sulfuric) acid and heavy metals (Goodell, 2006).

Distribution of coal occurs through the conventional fossil fuelled transportation infrastructure, dominantly (60%) via an extensive railroad system but also by barge and truck (Perry, 1983). For example, approximately 50 trains, each a mile or so long, leave the Powder River Basin daily, distributing its cargo around the country, a total of 220 million tons in 2004 (Goodell, 2006). Transport essentially triples the cost of coal to the power plant operator. Power plants typically store only a month's supply on site. The current U.S. railroad infrastructure is technologically outdated, underfunded and unsafe, as demonstrated by the many derailments reported in the media. Since the system is already overloaded, it presents a bottleneck for any future expansion of coal based power generation in the U.S. Urgent and massive investment is needed to modernize and expand the U.S. railroad infrastructure. However, in the U.S. this task clearly is technologically and economically achievable.

Coal quality is measured not only in carbon content but also in terms of undesirable contents of sulfur and toxic heavy metals. These impurities are of concern, since there is yet essentially no capture of the waste arising from operation of coal mines or power plants. Flue gases from coal combustion contain most of the sulfur, as well as nitric oxides from the burn process. Photochemical and wet-phase oxidation of nitric and sulfuric gases produces "acid rain" impacting biodiversity in water bodies and forests. The main metal contents of U.S. coal are mercury (0.01-33 ppm, Billings 1972), lead (~10 ppm, Chow 1972), uranium (~1 ppm, McBride, 1978, Gabbard, 1993), thorium (~ 3 ppm, Gabbard, 1993) and radium (Chernousenko, 1991), but also beryllium, chromium, manganese, and arsenic are present. Although these substances are of great environmental concern, even newer power plants routinely discharge a substantial fraction of the heavy-metal content of coal into the waste stream, e.g., vent them into the air or release them into rivers. Collectively, the coal-fueled power plants in the U.S. release annually 48 tons of mercury. They are also the major source of airborne uranium (40 tons annually) and radium. The heavy metal pollutants eventually precipitate and find their way into the human food supply.

Attempts have been made by governments, for example, by the U.S (Clean Air Acts), European countries, and China, to limit the emission of sulfur oxides and heavy metals by more or less strict regulation. Switching from the sulfur rich eastern coals to the lighter western coals containing fewer pollutants, some power companies were able to comply with regulations. As a positive result of this change in fuel and the installation of sulfur "scrubbers" in the flues of large power plants, the intensity of acid rain has somewhat diminished in the north-eastern U.S. However, overall legislation has been timid and its success has been varied. For example, substantial emission reductions for mercury are required mostly for new "clean" power plants and will likely take effect only over decade-long periods (EPA, 2005). As yet no specific plans have been published dealing with some of the other heavy metal pollutants such as thorium and uranium. At present, coal power plants in the U.S. emit at least 100 times more of these substances than the nuclear power and processing plants (Gabbard, 1993).

Of even greater health concern are particulate matter emissions from coal-fired power plants, as well as from automobiles. Particulate matter (PM) includes airborne dust, dirt, soot, smoke, and liquid droplets. Fine particles (< 2.5 μm in size) can be inhaled and lodge in the lungs (EPA, 2006). They cause respiratory and cardiovascular disease. PM pollution affects a very large fraction of the population, especially in large cities where the resulting smog and haze shorten life spans. Programs to reduce the PM pollution in the U.S. in the period from 1999 to 2006 effected a decrease of just 15% in PM output (EPA-PM2.5, 2006). It is objective of the DOE Strategic Plan (DOE, 2006) to essentially remove entirely PM (PM₂.₅) from coal power plant emissions by 2020. Whether this can be done without new and strictly enforced regulation is doubtful.

There are many R&D efforts in the U.S. and around the world to develop and test new "clean coal" burning technologies, which can solve some of the waste and pollution problems associated with fossil fuels (DOE, 2004). This is a prerequisite for coal to become again a main staple of energy economy. In addition to generating electricity, these technologies are also amenable to the production of hydrogen and synthetic liquid fuels, satisfying another basic requisite for a sustainable primary energy source. This capability is important in the pursuit of energy independence requiring flexible response in times of acute or persistent oil or gas import shortages. For example, in a coal gasification process developed by corporations such as Texaco and General Electric, coal, refinery tars, and other organic substances are, in the presence of oxygen, exposed to high heat and pressure and thereby partially gasified. Sulfur and other impurities can be removed from the coal gas, leaving "syngas", a mixture of $CO_2$, $CO$, and $H_2$, which in such plants drives a set of coupled turbine power generators.

This "Integrated Gasification Combined Cycle" (IGCC) has been used as design principle in several new coal power plants in the U.S and Europe (Wabash River, 1999, GE, 2001, NETL, 2002, Sarlux, 2007). The schematic flow diagram of an IGCC coal power plan given in Fig. 2 outlines the three main processes in coal refining: gasification under (separately produced) oxygen supply, gas component separation, and combined cycle combustion and steam reforming. In the latter scheme, the hot exhausts of a gas turbine drive a conventional steam turbine.

The IGCC plant is well suited for carbon capture. Carbon dioxide can be isolated and treated for onsite storage or transfer. For example, $CO_2$ in exhaust gases led through a monoethanolamine (MEA) solution forms a loosely bound chemical compound, from which it is released again by heating. The gas can also be reacted with magnesium oxide to form magnesium carbonate for storage. Most significantly, under high compression at low temperatures syngas is transformed into methanol ($CH_3OH$). Here, the $CO_2$ component combines with $H_2$ according to the (catalytic) reaction

$$CO_2 + 3H_2 \leftrightarrow CH_3OH + H_2O + \Delta H \qquad (2)$$

The enthalpy $\Delta H$ (=-9.8 kcal/mol) quantifies the energy release in this exothermic reaction, which is partially converted into electricity (at high process temperature $T_H$) but has partially to be taken out of the system by cooling (to an exhaust temperature $T_L$ as low as possible). The thermodynamic efficiency $\varepsilon$ of this (or any other thermal) energy conversion process is limited. It depends on the temperature spread $T_H$ -$T_L$ and cannot exceed

$$\varepsilon_{max} = (1 - T_L/T_H) \qquad (3)$$

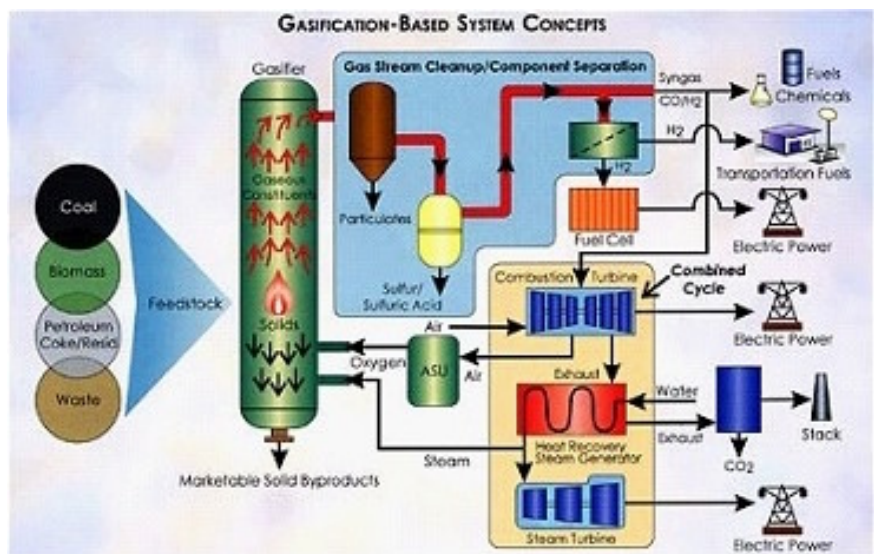

Figure 2: Schematic flow diagram of IGCC coal power plant with hydrogen and synthetic fuel production, as well as carbon dioxide capture (DOE, 2003).

Here, the temperatures $T_L$ and $T_H$ are measured on the absolute Kelvin scale (degree K = $^0$C + 273). For typical power plants, $\varepsilon = 0.3 - 0.4$ but $\varepsilon$ can exceed $0.6$ for a modern IGCC plant. Cooling requires ample supply of fresh water derived, e.g., from a nearby river. Hence the ubiquitous characteristic cooling towers. In "cogeneration," where the still relatively hot exhaust gases are used for residential heating, $T_L$ is lowered further and the overall process efficiency is increased. However, to take advantage of this efficiency requires proximity of large power plants to residential areas.

The carbon monoxide in the syngas produced in IGCC plants can be reacted with hydrogen, similar to Equ. (2), to produce the liquid fuel methanol. In addition, when reacted in $H_2$ excess, carbon oxides can be reformed to ammonia ($NH_3$), which is also a byproduct of oil refining. A number of other industrial processes of coal liquefaction have been used extensively in the past and are immediately available. They include the well-known Fischer-Tropsch process based on the schematic catalytic sum reaction

$$CO + 2H_2 \leftrightarrow (\text{--}CH_2\text{--}) + H_2O \qquad (4)$$

Here the parentheses indicate a mixture of several gaseous and liquid compounds, which have to be separated. Large scale applications of the process are presently running in South Africa (Sasol I-III) providing synthetic transportation fuels for the nation. More modern, simpler and more efficient processes can be used to produce methanol as liquid synfuel in plants that are part of, or coupled to, an IGCC plant. Importantly, Fig. 2 illustrates the complexity of an advanced IGCC power plant, which can take up to 6 years to construct and put online at a cost of the order of \$3B. For power companies, the design makes therefore direct economical sense only for fairly large plants with at least 0.8-1 GW$_e$ in power production. The total costs for capture and storage of a ton of $CO_2$ on site have been estimated to \$25, corresponding to some \$60 per ton of coal. This expense is expected to raise the production cost of coal electricity from 4¢/kWh$_e$ to 6¢/kWh$_e$. Such cost considerations are the main reasons why in the U.S. most construction permits, even for new coal power plants, have not been of the expensive IGCC type. Energy policies employing "carbon caps" or "carbon taxes and trading," may be required to redirect trends.

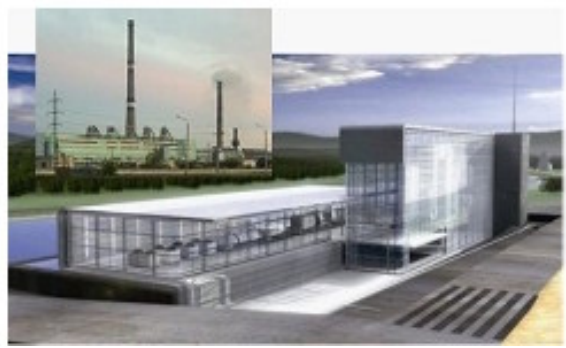

**Figure 3: Artist's view of a "FutureGen" IGCC clean coal power plant with $CO_2$ capture (DOE, 2004). The insert shows the photograph of a conventional coal power plant**

Motivated by concerns about the impact of anthropogenic GHG emissions on the global climate and biosphere, carbon capture and sequester (CCS) methods are investigated (NETL, 2005), by which $CO_2$ emitted by fossil fuel burning power plants could be deposited in permanent repositories. After all, carbon dioxide makes up for 60% of the GHG totals. Yearly, the U.S. alone produces 5.7 Gt of $CO_2$, or 20 tons per capita per year. To avoid $CO_2$ release into the atmosphere, the gas has to be isolated and captured, before it can be permanently deposited. The modern coal power plant envisioned in the FutureGen development program (DOE, 2004) features a closed gas/steam cycle and the onsite capture of all pollutants. As illustrated in the artist's view of such a modern coal power plant reproduced in Fig. 3, the most striking difference in appearance of IGCC plants is the ab-sence of tall smoke stacks, which are the hall-mark of conventional plants (insert in Fig.3).

The various carbon capturing sequestration and conversion schemes under study all require refined fuel treatment technology such as realized presently only in the very few new IGCC type coal power plants built in the U.S. and Europe. Prohibitively expensive technical retrofits, essentially reconstructions, would be required to allow at least $CO_2$ capture also from older "dirty" power plants. It should therefore become highest priority in a sustainable energy strategy to guarantee that all new coal power plants are licensed and built exclusively according to "clean coal" (e.g., IGCC) specifications. At a minimum, this would concern the construction of the more than 150 major plants (154 GW until 2030, EIA 2006) previewed in the U.S. To this number, an increasing number of conversions or replacements of conventional oil and gas fired plants has to be added.

Requiring permanent $CO_2$ sequestration adds another major complication to the process, and another major expense. Already the capture and storage of $CO_2$ at an IGCC plant would raise the cost of clean-coal generated electricity by approximately 50% and make coal economically less competitive for electricity production. The overall cost for the construction of a sufficiently large number of IGCC clean coal power plants in the U.S would be of the order of, even exceeding, 500 B$ over 20 years, requiring massive and sustained private and public investment. However, the prospect of coal as a basis for the production of liquid transportation fuel may turn out to be instrumental in gaining public support for a "clean coal" energy strategy.

In principle, $CO_2$ sequestration can be accomplished in various technological schemes. Biological carbon sinks can be created by planting trees or plankton algae farms which absorb atmospheric $CO_2$ in their photosynthesis. The practice would require enormous arable land or ocean surface areas and planting efforts that would go very much against the historical trends (deforestation).

Perhaps more promising geological sequestration involves pumping $CO_2$ into depleted oil and gas reservoirs, under the deep sea bed, in other geologically stable pockets such as in depleted coal seams, salt domes, or in porous sedimentary rock formations, where it could be safe for tens of millennia. Since oil reservoirs and coal seams are typically accompanied by natural gas or methane bubbles caught in anticlines of geological strata, geological sequestration of $CO_2$ is feasible in principle and can be done relatively safely. This is also supported by the fact that, in the U.S., some natural geological cavities are used extensively for temporary natural gas storage. Unfortunately, overall there are too few depleted oil wells or coal seams located conveniently in proximity of population centers serviced by coal power plants to make this method universally feasible without long-range transfer of the gas.

Sequestering the gas at the bottom of the deep ocean sea, the prevailing high pressure would form a "$CO_2$ lake." Since the ocean water is not saturated in dissolved $CO_2$, there is chemically little risk for the gas to bubble through to the surface and be released into the atmosphere.

Since 1996, a large-scale test of ocean sequestration has been underway some 250 km off the coast of Norway (Sleipner field). The project is schematically illustrated in Fig. 4. In this project, so far 8 Mt of $CO_2$ have been pumped into a porous Utsira sandstone layer 1,000 m under the bottom of the North Sea. The migration of the sequestered gas is monitored seismographically.

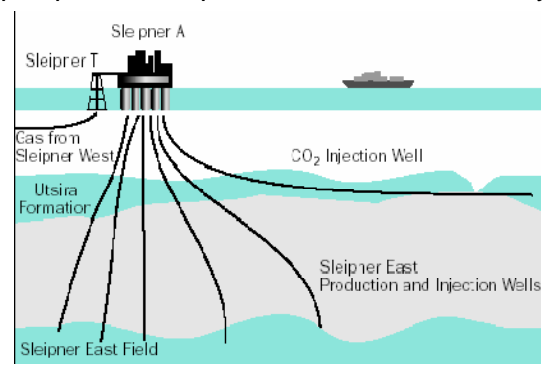

Figure 4: Schematics of StatOil's CO2 sequestration project in the North Sea off the coast of Norway (Statoil).

Obvious ecological challenges of carbon sequestration include the back diffusion of $CO_2$ to the surface and environmental pressures resulting from changes in pH value (acidity), for example, affecting maritime life. Sudden $CO_2$ release from deposits induced by earthquakes or volcanic eruptions poses potentially catastrophic hazards that are well known from the 1986 Cameroon Nyos Lake disaster. Here, 1700 lives were lost when such an eruption released suddenly large amounts of the heavier-than-air $CO_2$, suffocating the nearby village population. In case that $CO_2$ leakage from the Sleipner storage site into the sea should occur, the gas would be readily absorbed by the sea and constitute no harm to populations or sea lanes. This added safety feature does not exist in land sequestration, requiring continuous long-term monitoring of the sequestration status.

In principle, chemical sequestration is also possible, for example, by using or accelerating "weathering" reactions binding $CO_2$ to calcium compounds (Falkowski, 2000). Reacting $CO_2$ with lime (CaO) produces limestone (CaCO$_3$) suitable for temporary onsite storage. Heating limestone sets $CO_2$ free and regenerates the lime. From these and other potential processes considered, it is clear that carbon capture and sequestration (CCS) schemes would only be economical if performed on a large scale and for large (> 1 GW) power plant complexes, which would also have to be located far from population centers. To give an illustration of the scale, capturing and compressing $CO_2$ produced by existing U.S. coal power plants would require handling a volume of 50 million barrels per day (NPC, 2007), more than twice the oil volume moved in a day.

The U.S. DOE has provided a roadmap (DOE, 2004, NETL, 2005, DOE, 2006) to encourage and steer corresponding R&D in CCS. The objective of DOE's Clean Coal Program (DOE, 2006) is to demonstrate a zero-emission coal power plant by the year 2020. While carbon capture is technically feasible now and could be practiced, the viability of long-term carbon sequestration on dry land still has to be demonstrated, and enabling regulation has yet to be devised.

## B.  Oil and natural gas

Since the beginning of last century, inexpensive oil and, to a lesser extent, natural gas had displaced coal as a ubiquitous fuel and reshaped industrialized society. Oil products provide now about 40% of the energy consumed worldwide by industry and households, while gas and coal make up 25% each in that balance. In particular, crude oil derived petroleum, diesel, and kerosene fuels the transportation system in the industrialized world. In the U.S., about 65% of the crude oil serves that purpose. While the proven and likely coal reserves are substantial and could satisfy much of the world's energy demand expected for this century, such a statement cannot be made for crude oil. As shown in Fig. 5 ("Hubbert's Peak", GAO, 2007), the U.S. domestic oil production has peaked at about 10 million barrels of oil (bo) a day in the 1970s and has now declined to half that amount (NPC, 2007), while the increasing consumption has now exceeded 20 million barrels per day. An illustration of the decreasing prospect of domestic and global oil production is the fact that the major U.S. oil companies have not built a single new domestic oil refinery during the past 30 years. They are now even reducing the number of refineries they own, from the present number of app. 150, by selling to independent operators. The 10 Mbo/d of oil presently imported into the U.S. already occupy 2/3 of the throughput of the existing oil refineries, which are running at capacity (17 Mbo/d).

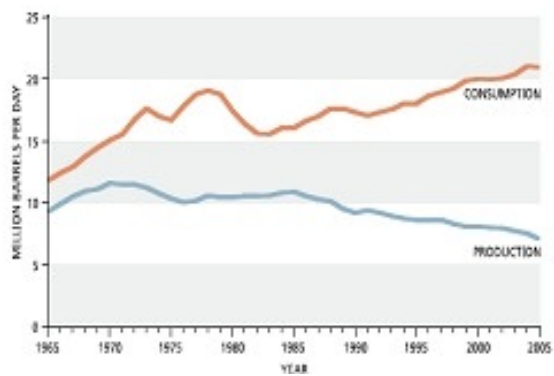

There are still sizeable reserves of recoverable oil, mostly (65%) in OPEC countries. Important are those in the Middle East, notably in Saudi Arabia, Iraq, Iran, Kuwait, and the United Arab Emirates. Other, somewhat smaller reserves are found in Russian Siberia, Africa (Libya, Nigeria, and Angola), and South America (Venezuela). In comparison, the much debated recoverable crude oil resources located in the U.S Arctic National Wildlife Refuge (ANWR) are miniscule, covering perhaps 4% of the U.S. consumption for a few years, 5-7 Gbo in total.

However, accurate assessment of oil reserves is not possible, and extrapolations of global oil production to the year of 2030 range from less than 80 to 140 million barrels a day (NPC, 2007). But accepting even such a low minimum reserve may turn out to be over optimistic. There is presently no universally accepted reporting standard, and the numbers for crude oil reserves quoted by different agencies diverge significantly. In addition, it is of great practical consequence that estimates are not differentiated with respect to the ease with which the oil reserves can be extracted. Clearly, crude oil is much more economical to produce from Saudi Arabia's land based oil wells than from below the seabed under the Gulf or the North Sea. In the Jack Field in the Gulf of Mexico, Chevron now drills 30,000 feet under the seabed to find crude in porous rock (Grisholm Little, 2007), hoping to produce perhaps as much as 5 Gbo. Plausibly, the now ubiquitous oil platforms are expensive to construct and time consuming to put into operation. For example, the Canadian Hibernia platform cost 5 B$ and took 19 years from discovery to production, while the Thunder Horse platform in the Gulf has cost so far 4 B$ and, eight years after discovery, has not yet taken up production (NPC, 2007). In addition to their high costs, oil platforms are susceptible to adverse weather conditions, as demonstrated by losses during the hurricanes Katrina and Rita in 2005, which caused instabilities in oil markets that were deemed unacceptable by the public.

Particularly difficult to produce is crude from the Canadian sticky tar "oil sands," a process requiring heat and large amounts of fresh water (McFarland, 2006). These latter reserves are so expensive to recover that some experts believe that they should be omitted from the roster of reserves. The remoteness of many of the oil fields from the consumers has led to coining the term "stranded oil."

Based on extensive data analysis, Deffeyes (Deffeyes, 2001) has predicted world oil production to peak already before 2008. Although that estimate may turn out to be slightly pessimistic, most expert studies agree that crude oil world production will have declined significantly within 15-30 years, well before mid century. The actual extent of the "cheap oil era" defining the modern industrial economies depends on multiple, uncertain factors, including possible improvements of recovery factors by new production techniques (horizontal drilling, steam or $CO_2$ forcing, etc.). Obviously, the future industrial growth rates, associated increase in global competition for crude oil and unfavorable political and investment conditions in countries with oil reserves will play major roles in shortening the cheap-oil era, even before the world reserves run out. Because of the dominating role of oil as a primary energy source, the consequences for the industrialized world will be substantial.

Compared to coal mining, the hazards of oil production are comparatively small. The distribution of oil from remote regions to refineries and consumers is considered relatively safe, and casualties are rare. Within the U.S. most of the transfer of crude oil and refined products is accomplished by a system of approximately 200,000 miles of a mainly underground pipelines, driven by powerful pumps every 50-70 miles. The Trans Alaska Pipeline System (TAPS) is a notable exception running mostly above ground. A network of 55,000 miles of large diameter (8"-34"dia., $1M/mile) crude oil trunk pipe lines is operated by companies like Shell, BP, Exxon Mobile, and others (see Fig. 6).

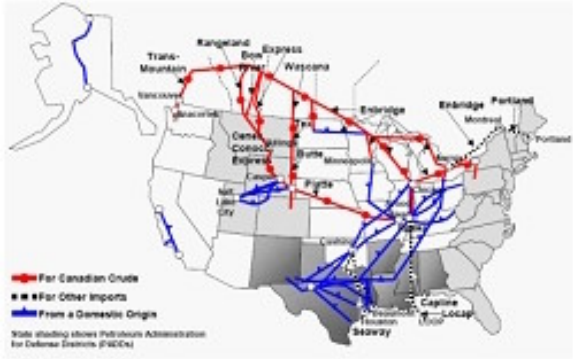

Pipelines are also used to move various refinery products (gasoline, diesel, heating oil, kerosene). The systems serve a large number of terminals with substantial storage capacity, functioning as hubs for local distribution. Only the latter task is dominantly accomplished by tanker trucks.

Nevertheless, fatal accidents have occurred in this energy sector and oil transport has been marred with the ecological disasters of numerous spills. In 1998, a massive oil explosion in the Niger Delta claimed more than five hundred lives (Darley, 2004). In addition, in 2001 the U.S. oil pipeline system suffered 129 spills larger than 50 barrels (2100 gal). Perhaps the most memorable oil spill is that of the EXXON Valdez in the Prince William Sound in 1989. In that accident, 11 million gallons of crude oil leaked out of the tanker, spoiled some 1,300 miles of the coast of Alaska, killed uncounted wildlife, and interrupted regional economic activities. Accident related costs amounted to several B$, and the cleanup took more than a decade to accomplish.

In spite of accidents and spills, the pipeline mode is regarded as the only technically and economically feasible distribution method. Because of its relatively low energy density (caloric value), the liquid volume that has to be moved is much too large for significant long-distance transport via rail, barge, or truck. If required by future demand and economy, the major pipeline network can be expanded with existing technology at the cost of approximately 1M$/mi.

Hazards associated with combustion are similar for all fossil fuels, as far as sulfuric and nitric gases or small particulates are concerned. Like coals, crude oils from different regions have different admixtures of sulfur. Because of the lower energy density of oil, however, specific GHG emission from oil combustion is enhanced (~10:7) as compared to coal. On the other hand, heavy metal content of oil is smaller than the mercury, thorium, and uranium content of coal. Hence, also a lesser amount of radioactivity is released in the combustion of oil than in coal burning. Nevertheless, the ammonia and mercury in the toxic waste stream from oil refineries led into rivers and lakes generates conditions adverse to aquatic life. So have recent expansion plans for an Indiana Marathon refinery been met with strong opposition from environmental and residential groups concerned about the already highly polluted Lake Superior (NPR, 2007).

Together with oil, natural gas makes up about 60% of the world's present primary energy consumption (NPC, 2007). The gas consists mainly of methane (85-95%), a highly potent GHG, some higher hydro-carbons, and nitrogen. The fact that in turbines and generators natural gas burns much cleaner and more efficiently than oil has motivated a widely spread switch from coal and oil to gas. This switch occurred somewhat later than generally expected (Smil, 2005) because its 1000-fold lower energy density (34 MJ/m$^3$) than oil makes gas more expensive to distribute and less suitable as a transportation fuel. Therefore, natural gas is now mostly produced for residential, commercial, and industrial consumption. Beside its role as primary energy source, natural gas is an important feedstock for chemicals, pharmaceuticals, fertilizer, fabrics and plastics.

Annual U.S. consumption in natural gas was app. 23 Tcf in 2004 (EIA, 2005), corresponding to 63 Bcf/day. The most recent data for production and consumption are depicted in Fig. 7 (NPC, 2007). The current trends are declining both in demand and production, which is partially attributable to the high cost of natural gas as compared to coal. Yet, other reports (DOE-NETL, 2005) predict again slight increases in U.S. domestic production and gains in natural gas consumption by 50% by 2025. According to some sources (EIA, 2006), approximately 130 GW$_e$ in new construction of gas-fired power plants are previewed until 2030.

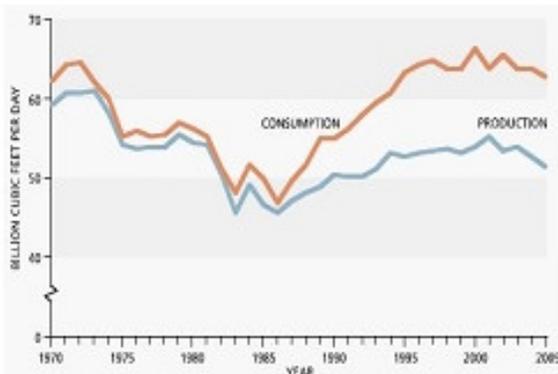

Naturally, the gas occurs often together with oil or coal (CBM, coal bed methane). Typically the gas is trapped in a bubble above an oil reservoir or in a coal bed. It can leak naturally or escape in the mining process, in either case adding to the atmospheric GHG repertory. Since it provides an explosion hazard at the production site, the gas is still mostly been "flared off" at an oil wellhead, i.e., just burned at the production site. At night, huge Russian, Iranian, Arab (Qatar), and Nigerian natural gas flares are easily seen from space pinpointing substantial gas resources. Other reserves are located in Trinidad/Tobago, the Asia/Pacific Basin (Indonesia), the South China Sea, Australia, and in the Arctic seabed. Similar to oil, the extent of the world's recoverable natural gas resources is subject to debate. Typical estimates (DOE-NETL, 2005) of gas reserves amount to 6,000 Tcf, corresponding to just 60 years of present global consumption. Given a rapid increase in global demand, Hubbert's peak for gas may occur long before mid-century, relegating natural gas to the status of a rather finite resource. Certainly, domestic gas production will be insufficient to satisfy the projected U.S. gas demand of 6-7 Tcf by 2025 (DOE-NETL, 2005).

Therefore, most of the gas for the planned power plants would have to be imported, likely in form of liquefied natural gas (LNG). Although expensive and vulnerable, the LNG delivery technique is suggested by the remoteness of natural gas resources. Liquefaction plants have to be constructed at the production site, typically at a cost of 1.5-2 B$ per plant. In future, alternative solutions may be used instead, such as currently tested in Qatar, where natural gas is converted to synthetic gasoline prior to shipment. Steam reforming of natural gas, as used with syngas in the Fischer-Tropsch process, together with partial oxidation at high temperature is an expensive but feasible method to convert gas into synthetic gasoline ("synfuel"). The method is being explored by industrial companies in several countries.

In the regular LNG process, the multiple liquefaction/revaporization cycles and the cryogenic cooling to -162$^0$C required during transfer are expensive and energy intensive. This and evaporative losses reduce the available energy content of the gas to app. 75% of its initial value. The process requires super-insulated storage and transfer vessels with multi-layered walls some 6' thick (see Fig. 8). Super tankers with capacities exceeding 200,000 tons are presently developed for LNG shipping. The energy content of such a ship resembles that effective in a small nuclear device. The lack of armor of these tankers and their unprotected exposure on the open sea lanes makes them soft targets. To an even larger extent, such a concern applies to the new, high-capacity LNG terminals.

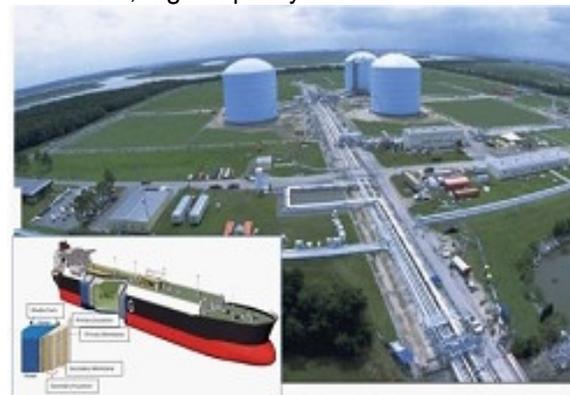

LNG operations in the past have not been accident free. After the earlier fatal LNG explosions of 1944 and 1973 in the U.S. (Darley, 2004), in 2004 several LNG trains exploded in Skikda/Algeria, claiming 31 lives and injuring 74. These repeated LNG accidents motivate serious security considerations of an "LNG economy," which would have per week a fleet of 100 or more LNG super tankers crossing the Atlantic, to be received and unloaded at U.S. terminals.

The U.S has currently only 6 LNG terminals, where LNG is received, regasified, stored short term, and pumped into an extensive system of presently 280,000 miles of gas pipe lines. Intermediate short-term LNG storage occurs at buffering ("peak-shaving") facilities, which scoop gas from the pipeline system and liquefy it for storage, to be able to revaporize and release the LNG again in times of peak demand. In the U.S, construction of 20 to 30 new LNG terminals is under consideration (DOE-NETL, 2005, Romm, 2005). It is noteworthy that the capacity of such a large number of LNG terminals would at most be sufficient to handle 50% of the projected imports. There are a number of reasons for this projected limitation. Certainly one of them is that the cost for construction and operation of an LNG chain, from production to delivery, is substantial (7-10 B$). However, other reasons, like uncertainty about future accessibility, safety, and profitability of LNG all play important roles. The actual rate of increase in U.S LNG consumption will also be subject to international competition, in particular from Japan, which is now the major LNG importer.

From the documented facts collected and listed above, which are all in the public domain, one has to conclude that among the fossil energy sources neither oil nor natural gas satisfies the requirements for a dependable primary energy source. In either case, the domestic peak production has been passed in the U.S. or is imminent, making the resource not abundant enough for a sustainable long-term energy economy. Similar observations hold for most other industrialized nations. Occasional minor new oil field discoveries could delay and mitigate the demise of the "cheap-oil economy" but cannot prevent it. To initiate a transition to another primary energy source is therefore urgent.

Countries in North America, Eurasia, Asia and South Africa, including the U.S., are in the fortunate position to have abundant and recoverable domestic energy resources in the form of coal. Here, coal generated fuels can provide a durable pillar and stability for the economy and displace oil and gas to the extent that these latter resources are depleted. However, sizeable investments have to be made to resolve the Strategic Issues for coal and make it a more benign, versatile, and acceptable primary energy source:

a) The hazards of mining have to be reduced significantly, e.g., by further introduction of remotely controlled robotics;

b) Restoration of ecology after open-pit mining has to be mandated (and made part of the price of doing business);

c) Emission of pollutants/GHG has to be reduced to acceptable levels, e.g., by CCS/IGCC "clean-coal" and CCS strategies; and

d) Petro-based transportation fuels should be increasingly replaced by synthetic fuels, achievable also with a new generation of IGCC plants.

These tasks are all achievable with existing technology but, of course, at a price. Good cost estimates of such a scenario are not available, but a rough extrapolation suggests that a doubling of the coal fraction in the U.S. energy mix, under the clean coal scenario, may require funding of the order of a trillion US$ over the next three decades. Since they concern infrastructure investments used by several future generations, an informed public will consider such efforts feasible and worthwhile.

**Potential and Risk Assessment:  Renewable Energy Sources**

The development and large-scale utilization of a variety of renewable primary energy sources should be an ultimate goal of any long-term energy science policy. In recent decades development of such technologies has been en-couraged and their use has been highly subsidized. In several countries corresponding R&D has been well funded and at increased levels. For example, in Germany renewable energy production has been subsidized both directly and by tax incentives. Energy producers using renewable technologies are paid rates needed to recover their investment in the long term (Morris, 2006). Thus renewable energy can cost up to US$ 0.15-0.20 per kWh, a factor of up to three higher than retail energy cost.

In order to make renewables economically viable and scalable, significant technological barriers have to be overcome. However, never is energy generation cost free or environmentally benign, as illustrated below for a few renewables.

Renewable energy resources face several *Strategic Issues* due to their
  a) Comparatively low energy densities and efficiencies;
  b) Intermittency, subject to local weather, diurnal or seasonal fluctuations;
  c) Substantial impact on local climate, ecosystems, and human habitat;
  d) Uncompetitive high price per produced kWh;
  e) Long ("pay-back") times for capital or energy investment recovery.

Customarily, the broader category of "renewable" energy resource encompasses hydro-electric power and the "green" energy sources of geo-thermal, wind and solar energy, as well as energy produced from biomass (wood, grass, etc.). At present, only hydro-power contributes significantly to the overall electrical energy generation in industrialized countries. The U.S. has 80 GW installed in conventional hydro-electrical power plants contributing some 8% to the electricity generated in the country, while the "green" energy sources fall at or below a total of 2%.

## A.  Hydroelectric power generation

Availability of hydropower is obviously contingent upon precipitation levels and naturally undergoes seasonal fluctuations. In particular during droughts or hot and dry summers, when electricity is at a premium, hydropower is less available. Hydro-electric power generation depends on the hydrodynamics of a relatively slow, laminar flow of water through generator driving water turbines. Therefore the efficiency of this mode of energy generation is relatively low, and hydro-power plants are often operated only as auxiliary facilities, not to provide base load.

Most of the growth potential of hydropower is in Asia, South America, and the territories of the former Soviet Union. In the U.S. and most industrialized countries, the growth potential of the hydro-electric power sector is rather limited. Since most large rivers have already been dammed, only few new plants could conceivably be built and operated economically. On the other hand, many dams and plants are nearing their useful design life of app. 50 years. Contributing rivers and reservoirs have accumulated sludge and need excavation and restoration. In particular, important questions of how eventually to decommission large dams have only been posed recently, and convincing answers have yet to be found.

In addition, great environmental concerns have been raised against construction of new hydro-power plants. The flooding of huge areas destroys not only forest and wildlife habitat. As a standing body of water, reservoirs develop an anaerobic and/or $H_2S$ toxic environment that is hostile to aquatic live. For example, the dissolved oxygen (DO) levels in Canadian reservoirs are so low that wood of the inundated forests is well preserved, making underwater logging of these forests profitable for companies. Other, tropic reservoirs are thought to contribute significantly to GHG emission (WCD, 2000).

In the flooding necessary to create reservoirs, human settlements have to be abandoned as well. More importantly, a number of dam ruptures have claimed numerous lives. To recall just a few, major recent accidents occurred 1979 in Morvi/India (30,000 fatalities), 1975 in Banqiao/China (86,000 fatalities), and 1972 in Logan/USA (450 fatalities). In more subtle and unpredictable ways, reshaping and re-routing of the natural river system change the ecology. Affected for example are runoff and sedimentation in river deltas with consequences for the associated coastal wetlands.

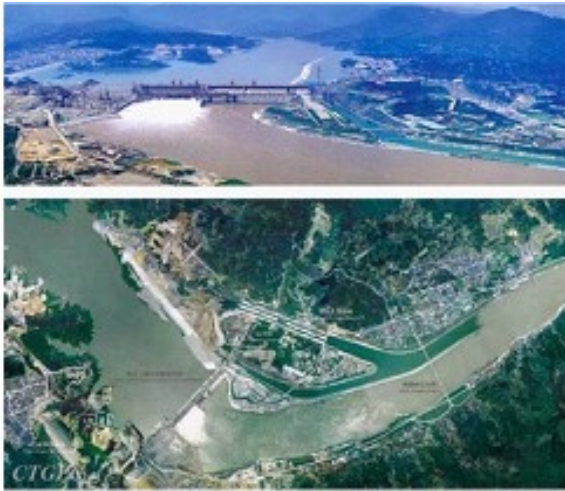

One of the exceptional new hydro-electric power plants is the gigantic, 25-B$, 18-GW Three Gorges Dam (TGD) project (Fig. 9) on the Chinese Yangtze River, scheduled to go online in 2009. The reservoir extends 600 km upstream from the 2-km long, 185-m high dam. Its construction has required the destruction of hundreds of villages and the displacement of an estimated 1.3 million persons. While the direct TGD construction costs should be recovered within 15 years of plant operation, the substantial collateral costs of the loss of arable land, fishery, and generally human livelihoods have not been estimated.

There are other, more exotic methods to extract energy from moving water bodies, such as from the ocean swell or the tidal waves. However, the economy of these methods has not yet been proven to be attractive. Hence, hydropower in its various present concepts is not expected to provide important answers to the future energy demand.

### B. Geothermal energy

In principle, geothermal energy is ubiquitous and virtually inexhaustible, because of the existing temperature profile of the planet. In practice, because of the generally weak temperature gradient in most locations, the economy of large-scale, renewable geothermal energy production is correlated with existing geological activity, such as volcanic or tectonic motion in the Earth's crust. The former is available, and heavily exploited, in Iceland, the latter along the Pacific Rim. Geothermal power plants targeting water in hot underground geothermal reservoirs also try to tap into a non-renewable energy resource. The efficiency of a more generic method of energy extraction, cycling surface water under pressure through hot dry bedrock at depths of several thousand feet faces technological hurdles which will be challenging and expensive to bypass.

The problems facing geothermal extraction technology are due to the fundamental thermodynamics of entropic heat flow. With presently only 3 GW produced, geothermal energy generation is marginal in the U.S. and will probably remain so, unless and until these fundamental barriers are overcome.

### C. Biomass energy production

Burning of biomass such as wood and animal manure (dung) is a relatively inefficient method of energy production ranging back to prehistoric times. More recently, biomass based methods have been proposed to produce biodiesel and ethanol-type fuels for the transportation sector. In WWII and post-WWII Europe, wood gas, a rather inefficient fuel, was used temporarily in the transportation sector.

The method of deriving energy from annual crops, rather than from forests or fossil source, is hampered by low energy yield, the need for large amounts of fertilizer, and the vulnerability of crops to weather and drought conditions. In contrast to its fossil equivalent, the small energy yield of 50 W/km$^2$ (Andrews & Jelley, 2007) obtained with the present biomass technology requires unacceptably large arable land areas and massive amounts of nitrous fertilizer to produce significant amounts of fuel. Effects of pollution generated by over fertilization and of long-term agricultural monoculture drive environmental concerns about biomass energy technology. It also appears ethical and untena-ble to use food for transportation fuel rather than for human sustenance. In addition, such policy will hasten the end of a cheap-food era.

An important argument made in favor of fuels derived from biomass is their $CO_2$ neutrality, although this argument does not extend to nitrous gases, which are also produced in biomass combustion. As base for such biofuels, one uses sugar cane in South America, rapeseed in Europe, and corn in the U.S. These synthetic fuels are blended with gasoline in different proportions. By 2010, European countries are claiming to have a fraction of approximately 6% of the consumed fuel generated from biomass (Morris, 2006).

However, even such modest predictions are viewed by many as too optimistic. For example, it appears debatable even, whether or not the production of ethanol from corn and other biomass has a positive or negative energy balance. A Cornell/Berkeley study (Pimentel, 2005) reporting negative energy balance for the production of ethanol from various plants has been contradicted by others (NBB, 2005). The question has to be considered as open.

Currently, the process biomass/fuel conversion is profitable to the agro-industry mainly because of farming subsidies granted by the government. In the U.S. ethanol production uses currently some 14% of the corn stocks (USDA, 2006). In most countries this type of land use does, or would, conflict and compete with food production. Already at this early stage of biomass utilization, rising corn prices have been reported (DoA, 2007, Brainard, 2007) in the U.S. attributed directly to the new use of the plant for bio-fuel production.

The core barrier in exploiting the more abundant energy source inherent in the cellulosic biomass consists in the difficulties associated with biological and chemical degradation of the material that eventually leads to ethanol production. A research initiative has been advanced by the U.S. Department of Energy (DOE-BIOM, 2006) to find energy richer cellulosic crops and more efficient biomass/fuel conversion processes. To judge the further prospects of biomass for energy generation will have to await R&D expected to become available by 2015-2020. The expectations are at best guardedly optimistic.

### D.  Wind energy

Wind power is currently the most important of the alternative ("green") energy technologies. Similar to hydro-mechanically driven mills and simple machines, the effects of mechanical drag and lift produced by aerodynamic flow have been utilized to generate energy (work) since centuries. The tall wind towers housing powerful electric wind turbines that seem to spring up in many locations, mainly around Europe, are simply modern realizations of 13[th] and 14[th] century European rotating-tower, horizontal-axis windmills. They convert ubiquitous me-chanical wind energy directly into electricity, without the detour via thermal plants. In the U.S., early wind farms have been installed at the Altamont Pass in California with over 5,000 relatively small wind towers (Fig. 10, top).

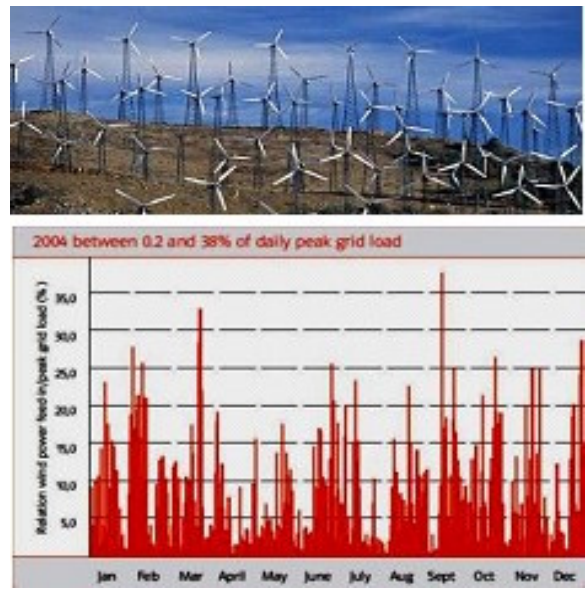

Figure 10: Partial view of wind farm (California) (Top). Efficiency recorded in 2004 for the electric power generation of a large scale assembly of wind farms (E.ON, 2005)

To utilize the higher wind speeds at altitude, the newest designs feature towers up to 100-150 m tall and have 2-or 3-blade propellers with diameters of more than $d$ = 100 m, which rotate with 10-30 rpm, corresponding to blade tip speeds of up to 130 mph. Advanced turbines generate up to 5 MW but more typical is 1-3 MW. The popular Vesta V90 turbine plant has a total height of 150 m to the blade tip and is rated at 3 $MW_e$. Wind turbines have an operational range of useful wind speeds (3-25 m/s); they are unusable in either low- or high-wind conditions. Propellers blades have to be "feathered" in high winds to avoid damage. Aerodynamics dictates minimum distances of $r = (5-10) \cdot d$ between the individual units of a "wind farm" (Sørensen, 2000), requiring an area of app. $10^5 m^2/MW$. A modern 500-unit wind farm producing nominally a power of 1 GW requires thus the substantial land area of 100 $km^2$.

Wind farms, often touted as the most environmentally friendly mode of energy generation, have nevertheless met with increasing resistance by the public. Concerns, which are difficult to quantify, range from the negative esthetic impact of wind farms on scenery to disturbing propeller noise. However, concrete environmental hazards created by these plants have been demonstrated by the observed average fatality rates per turbine and year of 2.3 birds and 3.4 bats, which have been reported for the U.S. in 2004 (NWCC, 2004). In addition, a large number of insects is struck by wind mill propellers. It is currently not known to what extent the avian mortality signals an emerging, deeper, and perhaps more significant, long-term ecological consequence of wind power.

The dissatisfying economics of wind power are more immediately and objectively accessible. They have been studied and reported (E.ON, 2005) for Germany, which has become the global leader in the use of this technology. In 2004, the country operated wind farms rated at a total of 16.4 $GW_e$. The E.ON Group controlling 7 GW of that capacity describes some of the issues faced in large-scale field operation:

- Wind power feeds energy into the grid only at a fraction of the rated power of the plants.
- In principle, wind power in present technology can replace conventional power plants only to a limited extent.

- Predictions of wind power feed into the grid are highly inaccurate.
- A new electric grid infrastructure is needed for wind power.

An important, illustrative data set is provided in Fig. 10 (bottom panel), showing the percentage power feed into the German electric grid by the E.ON group's 7-GW extended ensemble of wind farms. The actual numbers fluctuate wildly between 0.2% and 38% in 2004 (E.ON, 2005).

Both intermittency and low yearly average has led industry analysts to the estimate that by 2020, even under optimistic conditions of 40 GW rated wind power installed, only 4% of conventional power plants can nominally be replaced by wind power (E.ON, 2005). Even more significantly (and memorable), for every 1 GW of new wind power generators, new power plants of 1 GW of traditional technology would have to be put online (E.ON, 2005). Continuously running conventional (fossil fuel and/or nuclear) power plants with a stand-by capacity equal to the maximum wind power are required. This odd relationship could be remedied by future new grid and energy storage technologies. In the interim, a new very-high voltage electric power grid is to be built, which would accept and dampen associated fluctuations.

Accepting even the optimistic predictions by the interested, highly subsidized industry, one concludes that for the next few decades wind power is not likely to play a major role in the base load energy portfolio of industrialized nations. As overall power demand increases, the fraction of wind power will even diminish (E.ON, 2005). Specifically, wind power cannot be expected to provide reliably the base power load on which an economy has to rely. Future breakthroughs in energy storage or grid technologies could change the prospects of wind power.

### E.   Solar energy

The continuous direct influx of solar radiation energy ("insolation"= 1 kW/$m^2$) on the surface of the earth is many orders of magnitude larger than what is needed to sustain human energy demand. This fact has inspired the im-agination of the public and has generated probably unrealistic expectations of the ease with which future energy demand could be satisfied at a sustainable level. Unfortunately, even after decades of generous funding of the solar technology sector, starting in the 1960s with NASA's space program, technical/economic problems associated with conversion, capture and storage of sun radiation have remained largely unresolved. Correspondingly, in the U.S. presently only 0.02% of the electricity is generated directly from solar radiation.

The most popular technology employs photovoltaic (PV) silicon (Si) cells converting radiation directly into electricity. PV cells are arranged in solar panels. They are constructed as thin layers of "doped" Si (300 $\mu$m), window layers, and metallic contact films on a glass substrate. The doping, diffusing into the Si lattice atoms of electron donating and accepting materials, produces a diode device, which in its "dark" state has no free charge carriers (electrons or holes). Photons within a narrow band of wave lengths (energies) of the solar spectrum can be absorbed by the Si diode by promoting electrons from the lower (valence) band of electronic states to the higher (conduction) band, which produces a net electric current that can drive a load. Photons of higher than the characteristic Si-gap energy are ineffective, lead to a heating of the crystal lattice and a deterioration of its effective qualities. The overall conversion efficiency of a Si PV cell is a combination of surface reflectivity, absorption probability and internal power conversion probability.

As shown in Fig. 11 plotting efficiency vs. areal cost of PV cells, there are hard physical limits to the theoretical conversion efficiency for solar radiation. The currently accessible domain is labeled as Region I in this figure. Since the Si crystal material used for the production of the most efficient Si PV cells is a byproduct of computer chip production, PV unit costs have benefited from the efficiency of the computer industry.

The production of Si wafers occurs currently in clean laboratories and factories, which limit environmental impact. Nevertheless, the process requires handling of chemical toxins such as concentrated acids used in surface preparation (etching). A future production of PV cells in numbers large enough to satisfy a significant fraction of the energy demand would require handling and deposition of enormous quantities (many thousand tons) of these toxic chemicals, not only acids but also poisonous arsenic, selenium, indium, and other toxic materials (Eerkens, 2006). The environmental problems associated with PV production would grow to proportions characteristic of large-scale chemical industry.

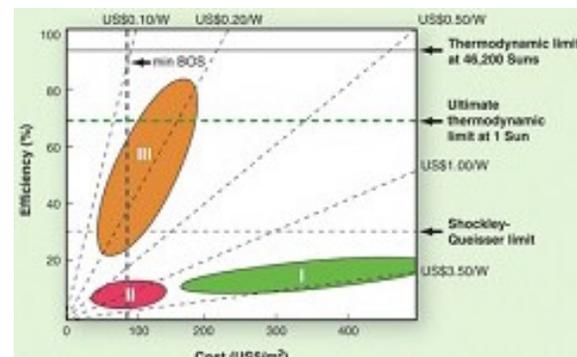

**Figure 11: Efficiency of photo voltaic cells vs. areal cost (Lewis, 2007).**

Currently the efficiency of direct energy conversion by commercial photo-voltaic solar cells is of the order of 15%, which is insufficient to make the technology economically viable. As an illustrative example, the integrated solar insolation of 1GWh/m$^2$ typical for Northern Europe or America, combined with the 15% efficiency of PV cells, translates to a specific energy conversion density of 150kWh/m$^2$. To displace the 7 TWh of energy produced yearly by one conventional 1-GW power plant requires an effective are of 46 km$^2$ covered with *Si PV* panels. This array of solar panels would presently cost 20-30 B\$. Such an amount would pay for 5-10 nuclear power plants, each of similar capacity (1 GW).

In addition, while the PV efficiency is initially close to 15%, the exposure of cells to radiation, weather and other interactions (bird droppings, sand blasting and deposition) causes lattice damage and surface erosion. These effects degrade PV cell performance significantly after prolonged use in the open field, for example in a sand desert (Smil, 2005). In practice, the useful life of solar panels may be much shorter than the 30 years often assumed.

New R&D is underway exploring ways to enhance efficiency and handling of solar cells. More complex, multi-layered, multi-well PV cells would absorb and convert a larger portion of the solar spectrum. Semiconductor materials such as GaAs, CdS, or CdTe are more efficient than Si, which has poor light absorbance. But these materials are also more expensive to produce. Other routes include cheaper amorphous silicon, rather than single-crystals. In principle, it is possible to emulate the electronic band structure that is absent from amorphous Si by chemical treatment. However, stable semiconductor diodes have not yet been produced by such technology.

Similarly, higher efficiencies could be reached with thin-film, flexible PV cells, if efficient light trapping technologies can be developed. Presently, their efficiency is even smaller than that of conventional Si photovoltaic cells.

Some hope to overcome theoretical (e.g., Shockley-Queisser) efficiency limits (Fig. 11) by developing new semi-conducting materials (PbS) and utilizing mesoscopic structures known as "quantum dots" or "nano-rods." These are examples of strategies by which one may eventually overcome technological barriers to the production of the efficient devices in Region III (Fig. 11) required for large-scale applications.

In addition to PV, and following a less efficient route, field experience is being gathered with several thermal solar power plants in the U.S. and Europe. Examples are the Solar One/Two plant in the Mojave Desert near Barstow/CA (Fanchi, 2006) and a new plant in Almeria/Spain. Here solar radiation is intercepted by helio-static mirrors and focused on a receiver atop a tower. The steam generated is used to drive a turbine. Variations of the principle use molten nitrate salt as heat transfer and storage fluid.

There are no clear indications that the efficiency of this latter technology could be raised much beyond its presently achieved maximum of 15%. Intermittency, energy storage, and grid feed-in problems discussed earlier for other methods hamper this and all solar energy technologies. In summary, solar energy technologies need to be improved significantly, before they can be considered a reliable primary energy resource. However, in principle, solar energy generation appears to have a great potential and poses new and scientifically interesting questions, justifying sustained R&D support.

In summary, one has to conclude that there is currently no technology available to resolve the main Strategic Issues afflicting renewable energy, although certain avenues promise progress in future decades. These conclusions about renewables, together with the realization that oil and gas reserves are finite and that their exploitation may adversely influence the global climate, have rekindled interest in nuclear power as a viable, long-term primary energy technology with a small "carbon footprint." Arguments have been given previously for the utilization of remaining coal reserves for purposes other than electricity generation and a corresponding ramping up of nuclear power for electricity.

### Potential and Risk Assessment: Nuclear Energy

Nuclear transmutation is the most energetic process known. It is also ubiquitous, it fuels stars like the Sun, drives the evolution of the universe, and has numerous applications, from energy production to diagnostics and therapy in medicine. The process is essentially a rearrangement of the nucleons of a system of one or two initial atomic nuclei into the final product nuclei which have different average bonding strengths between the constituent nucleons (neutrons and protons) and can carry intrinsic excitation,

$$A_1(N_1,Z_1) + A_2(N_2,Z_2) \rightarrow \ldots \qquad (5)$$
$$\ldots \rightarrow A_3(N_3,Z_3) + A_4(N_4,Z_4) + Q$$

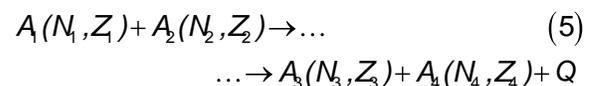

Here, $A_i$, $N_i$, and $Z_i$ are mass, neutron, and atomic numbers of the respective nuclei. Intermediate, transition-state nuclei have been left unspecified in Equ. 5. The quantity $Q$ is the energy release in the process, which is exothermic ($Q > 0$), if the products of the reaction (5) are more tightly bound than the initial nuclei.

The average binding energy per nucleon ($B/A$) is plotted in Fig. 12 vs. the mass number $A$ of each nucleus. Since iron (Fe) and nickel (Ni) are the most tightly bound, most stable nuclei, both aggregation (fusion) and disintegration (fission, spallation) release substantial amounts of energy when they proceed into the respective directions nearer to iron/nickel-like product nuclei. Therefore, both types of nuclear process can be utilized to generate energy. The interesting but technologically extremely challenging route of energy production by the fusion reaction of deuterium ($^2$H) with tritium ($^3$H) will not be discussed here.

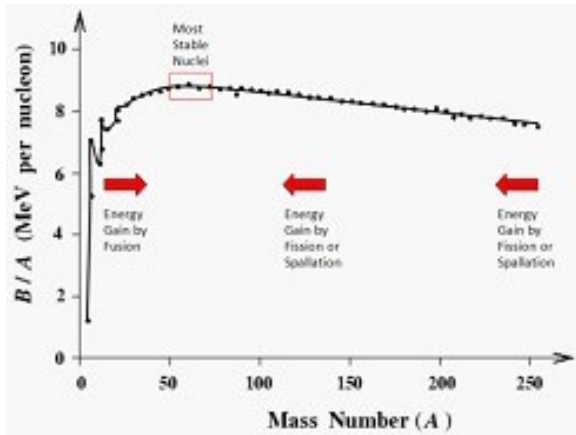

**Figure 12: The specific nuclear binding energy B/A per nucleon vs. mass number (A). The box indicates the most stable Fe/Ni region.**

The most common process exploited in nuclear power reactors is fission of uranium induced by "thermal" neutrons. Here, a low-energy ($E_n \sim 25$ meV) neutron is captured by a nucleus of the rare (0.7% abundance) isotope $^{235}$U, producing an excited $^{236}$U nucleus. The latter then splits into two massive fission fragments and 2 or 3 energetic (MeV) neutrons. A typical fission reaction is

$$n_{th} + {}^{235}U \rightarrow {}^{236}U^* \rightarrow {}^{147}La + {}^{87}Br + 2n + Q \quad (6)$$

where the energy gain is $Q \sim 170$ MeV. Each of the two new fission neutrons is "fast." If slowed down sufficiently in a moderator medium, they can in turn be captured with very high probability by a $^{235}$U nucleus and induce secondary fission reactions. Moderating medium in pressurized-water (PWR) or boiling-water (BWR) reactors is typically light water ($H_2O$), but heavy water ($D_2O$) or graphite is also used. The former ($D_2O$) moderator is hallmark of an intelligently designed Canadian reactor type (CANDU), which runs on natural uranium.

A fission chain reaction is generated and sustained in a reactor, if per fission reaction at least one ($k \gtrsim 1$) neutron remains within the reactor core, is "thermalized" by the moderator and available for capture. Introducing materials like cadmium (Cd) into the reactor makes it "sub-critical," diverting neutrons from being captured by uranium, which extinguishes the fission chain reaction ($k < 1$). By partially inserting Cd rods into the core, the fission chain reaction can be controlled and finely tuned. The buildup of fission products in the reactor core and the disappearance of fissile $^{235}$U have a similar effect, eventually extinguishing the chain reaction automatically. This requires refueling with fresh or reprocessed fuel.

The more abundant (99.3%) $^{238}$U isotope can also capture thermal neutrons but does not directly fission in the process. Therefore, conventional fission reactors use fuels enriched to app. 5% in $^{235}$U. Fission of $^{238}$U can be induced by fast-spectrum neutrons with energies higher than $E_n = 1$MeV. On the other hand, thermal-neutron capture by the "fertile" isotope $^{238}$U leads to a "breeding" of the fissile isotope $^{239}$Pu, via the chain

$$n_{th} + {}^{238}U \rightarrow {}^{239}U \xrightarrow{\beta-decay} {}^{239}Pu \quad (7)$$

In this fashion, the burning of $^{235}$U fuel in a reactor produces more nuclear fuel, the highly fissile plutonium $^{239}$Pu and, by subsequent neutron capture, also the non-fissile $^{240}$Pu. However, it may take years with such a U-Pu breeding process to produce amounts sufficient to load a reactor. The plutonium $^{239}$Pu fuel "bred" early by the above process from relatively fresh fuel could *chemically* be separated and in principle be used to produce high-yield nuclear weapons. Longer (>1 year) residence of the fuel in the reactor core produces an isotope mix unsuitable for weapons. Materials produced by a breeder reactor running on Th/U (oxides) fuel are even less amenable to such diversion. Here $^{232}$Th is the initial fertile isotope, and $^{233}$U is the fissile product isotope. Small admixtures of $^{nat}$U to the main Th component render the uranium isotopic mix unsuitable for nuclear weapons but also produce small amounts of $^{239}$Pu. If exposed to thermal neutrons for sufficiently long time, long-lived transactinide isotopes are bred from the initial reactor load, for example, neptunium (Np), americium (Am), and curium (Cm). The Th/U cycle also works with fast neutrons.

Albeit with lower probabilities, reactions induced by high-energy (~1-GeV) protons, delivered to the reactor core by an external accelerator, also induce fission of heavy nuclei like lead (Pb) and bismuth (Bi), thorium (Th), all isotopes of uranium and other actinides (Tishchenko, 2005). For example, fission reactions of the type

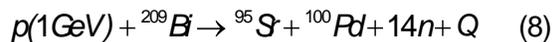

$$p(1GeV) + {}^{209}Bi \rightarrow {}^{95}Sr + {}^{100}Pd + 14n + Q \quad (8)$$

occur, but with probabilities of only 10-25%, compared to ≥95% for thermal-neutron induced fission of $^{235}$U or $^{239}$Pu. However, a number of

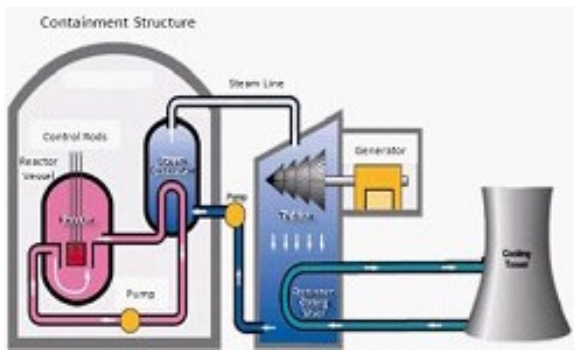

**Figure 13: Schematic of a nuclear fission reactor. The containment building houses reactor, primary cooling loop and steam generator.**

other processes such as

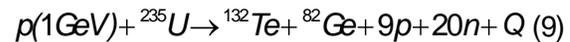

$$p(1GeV) + {}^{235}U \rightarrow {}^{132}Te + {}^{82}Ge + 9p + 20n + Q \quad (9)$$

involving many energetic neutrons and other light particles make up a larger fraction of the total reaction probability. These fast particles produce further exothermic reactions in the reactor and help sustaining a chain reaction.

Such accelerator driven nuclear reactor systems (ADS) could be considered the nuclear equivalent of a *Diesel* engine, which is able to produce energy from a multitude of fuels, including "spent" nuclear fuel elements ("waste"). Although a proof of principle has already been given (Andriamonge, 1995), other ADS test experiments are ongoing. *Regardless of their detailed outcome, immediate applicability of the ADS concept with uranium (or thorium/uranium) reactors is based on the important fact that per spallation reaction on average a large number (<$M_n$> ~ 25-30 for Bi-U) of fast neutrons plus a significant number of other light particles (p, d, t, He,..) is emitted (Herbach, 2006). Without the external particle beam, reactions in the subcritical (k~ 0.9) reactor do not take place. The large neutron multiplicity can also be used effectively in subcritical breeder and transmutation reactors discussed further below. The technology is essentially ready to be used.

The energy balance *Q* in a conventional reactor is highly positive: The fission of the nuclei contained in a single gram of uranium reactor fuel (3.5% enriched in $^{236}$U), into two massive fission fragments and 2-3 neutrons produces as much energy as the burning of 200 *kg* of coal, or 10,000 *L* of crude oil, or 73 *kg* of natural gas (LNG). The nuclear process energy is set free in form of kinetic energy of fission fragments, other charged particles (alpha particles), and prompt and delayed neutrons. Another 20 MeV of energy is delivered as electrons ("betas") and in form of electromagnetic radiation ($\gamma$-rays and X rays) in a "delayed" fashion through the beta decay of the fission products.

Like other power plants discussed earlier, a modern nuclear reactor is still basically a classic heat engine with a Carnot-type efficiency de-pending on the difference between operating and exhaust temperatures (cf. Equ. 3). Since the fission products dissipate their kinetic energy within the reactor core, the core heats up. The heat generated in the process is transferred by a cooling medium to a primary heat exchanger and to a steam generator. The steam is directed by pipes to the outside, where it drives a turbine and electric generator. Spent fuel contains, as radioactive "waste," long-lived fission products (FP) and newly formed actinide materials, which still contain recoverable nuclear energy that remains unused in current reactors.

Figure 13 shows the schematics of a pressurized water cooled nuclear (PWR) reactor. For safety reasons, in all Western designs, the reactor vessel, the primary cooling loop, and the steam generator are housed in a primary containment surrounded by a larger, secondary containment building, a massive structure made of several-feet thick armored concrete. The Russian RBMK-1000 reactor at Chernobyl did not have such a containment structure. However, following the accident, Russian nuclear plants have also been upgraded accordingly.

The scheme of energy generation from thermal-neutron induced fission of uranium described above is used by all 443 commercial nuclear power reactors operating presently in 31 countries around the world at a power output of 365 GW. Thorium is going to be the fuel for a new generation of reactors introduced in India. The number of countries considering nuclear power for electricity is growing, and an estimated 1,000 reactors are expected (Deutch, 2003) to be on grid by mid century. Of the order of 50 new reactors are now in the planning stage, 18 of them to be constructed in the U.S. For example, in June 2007, the Toshiba Corporation received an order for two 1.3-GW advanced boiling water reactors to be built in Houston, TX, for a total of US$ 4.86 B (CNN, 2007). However, most of the nuclear plants were built in the 1980s and 90s. Over the past decade, their efficiency has reached well over 90% of capacity. The 100 $GW_e$ of power generated by the 104 U.S. nuclear reactors accounts for 20% of the electricity consumed in the coun-

try. The corresponding number for France is 78%.

Susceptibility by the public to popular arguments against nuclear power is largely influenced by the perceived similarity to nuclear weapons and by the prevailing depth of understanding radiation. It is generally not fully appreciated that natural radioactivity has an important role to play in the evolution of life in the universe. The process releasing nuclear energy occurs also in nature as a result of the decay of unstable nuclei that have either been produced by cosmic rays in the atmosphere, or contained in rock, water, as well as in biological materials such as human tissue, bones, body fat, etc. The quantum mechanical tunnel effect is responsible for the sometimes extremely long delays ("life times") after which unstable nuclei decay "spontaneously" by emitting ionizing radiation.

On the other hand, the existence of adverse effects of ionizing radiation on biological tissue has been widely publicized, first during the 1950s, in the context of nuclear weapons. Major biological effects of radiation are similar to those of many chemicals. Both produce in the cells of biological tissues highly reactive organic molecules (radicals) that can attack cell DNA, induce mutations and, in the limit of high doses, cause cell death. However, unlike chemicals, particle and gamma radiation can be highly penetrating and damaging to tissue because of energy dissipation at depth.

Human radiation exposure is 85% due to natural sources (Lilley, 2001), with medical diagnostics and treatments causing most of the remainder (14%). Fallout from the 1950s bomb tests and the nuclear industry contribute a small fraction of one percent to the average dose. As pointed out previously, more significant than the latter and releases from the nuclear industry are amounts of airborne radioactive materials emitted in the combustion of coal.

Nevertheless, for a large part of the 20[th] century, public attitude toward nuclear power has been positive. Changes during the 1970s and 80s were induced by reactor incidents and political activism, as well as supported by competing commercial interest. In 1979, malfunction of one of the reactors at the U.S. Three Mile Island (TMI) nuclear plant led to a partial core melt-down. Radioactivity was safely captured by the double-shell containment building, and no one was harmed. A reactor accident at Chernobyl/Ukraine in 1986 resulted in 45 fatalities and displaced some 116,000 people living in a 30-km radius of the reactor. It also produced a radioactive fallout plume over Europe, but fortunately caused no further proven fatalities, although some studies have predicted illness, reduction in life quality and life expectancy for thousands more (Chernousenko, 1991, WHO, 1999). It is important to emphasize that an accident of this severity was never physically possible with reactors of Western design.

These events contributed to an erosion of the previously strong public support for nuclear power in the U.S. and several European countries. But, as a positive consequence of the unfortunate accidents, reactor safety requirements were increased to an extent that the safety record of nuclear power plants now represents a stellar example for all industries. On the other hand, to obtain the several individual licenses needed for construction, commissioning, and operation of power plants was turned into a difficult, costly, and protracted bureaucratic process whose eventual outcome was not always predictable. These regulatory facts contributed to long planning/construction periods and large cost overruns for nuclear power plants. More importantly, the global glut of cheap oil and gas during the past decades has made nuclear power, like coal, economically less competitive and less attractive.

To effectively develop and expand the nuclear energy option in the mid to long term, and to an extent that will enjoy public acceptance and support, several *Strategic Issues* associated with the technology have to be addressed. These issues include environmental hazards but are not limited to them. Concerns voiced in public discourse relate to conventional nuclear power using thermal-neutron induced fission of [235]U, specifically

A.    Operational reactor safety;
B.    Resource limits of nuclear fuel ([235]U/Pu);
C.    Safe capture, storage and (eventual) sequestration of radiotoxic nuclear waste;
D.    Prevention of proliferation of nuclear materials that can be used for weapons;
E.    Economy of nuclear energy.

The discussion below attempts to summarize the set of presently available answers to these strategic issues and give an outlook to the future discovery potential. The discussion will demonstrate that the issues have already been, or can be, resolved in present technology or in technology "around the corner." A positive role is played by a new Global Nuclear Energy Partnership (GNEP, 2006). The international GNEP effort includes the U.S., but not as the main technological driving force, a role that the U.S. has given up voluntarily in the 1970s and 80s. GNEP will promote standardized, modular, advanced reactor designs minimizing radioactive waste and the development of safe fuel cycles. Several of the new reactor alternatives are based on well understood science and engineering and none appear to face technological barriers.

There is still important engineering R&D to be accomplished and remaining challenges are left to basic nuclear science (physics and chemistry). The U.S. Nuclear Energy Research Institute (NERI) established in 1999 has been engaged mainly in advancing nuclear reactor technology in the U.S. But in principle, it supports also fundamental nuclear science. Currently 36 R&D projects by nuclear engineering departments are sponsored by NERI (DOE-NERI, 2006).

### A.    Reactor safety

As indicated previously, following the TMI incident and the Chernobyl accident, stricter regulations were imposed in the U.S. and elsewhere on reactor design, reactor operational safety, and operations management. The causes for these easily preventable reactor accidents have been investigated and are well understood. Underlying design flaws have been remedied by modification of existing plants and have been avoided in new designs for all now existing or planned reactors for the U.S., Europe, and Asia. The maximum credible accident (MCA) admitted by the new rules is much less severe than even a partial, TMI-type core meltdown with no radiation release to the outside. With the experience of the most recent 8,000 worldwide reactor-years, the nuclear power industry is now recognized as the safest of all major industries. These measures have rendered Strategic Issue A in the above list of five such issues as resolved. Nevertheless, future nuclear reactors will intrinsically be even safer. They will feature fail-safe operation through a negative reactivity factor, a passive emergency cooling system, advanced fuel element designs, etc.

The evolution in time of nuclear reactors is illustrated in Fig. 14 in terms of design categories defined by the U.S. DOE. Many of the earliest Generation I (Gen I) plants are still in operation, while the bulk of the presently operating (Gen II) light-water reactors (LWR) were built in the 1980s and 90s. Generation III reactors are conceptually similar to Gen. II reactors but feature enhanced safety. Some countries such as South Africa and China are pursuing more ad-

thoroughly studied, modelled (computationally simulated) and tested:

- Very high-temperature reactors (VHTR)
- Sodium-cooled fast reactors (SFR)
- Reactors cooled by (thermodynamically) supercritical water (SCWR)
- Lead-cooled fast reactors (LFR)
- Gas-cooled fast reactors (GFR)
- Molten-salt reactors (MSR)

These new reactor types promise more reliable operation, minimization of accident consequences, and improved accident management on site. It is plausible that these goals will be achieved by smart, inherent safety features that rely only on natural forces (e.g., gravity) and "can be understood by nonexperts" (DOE-GIF, 2002). These features include a robust and reliable control of reactivity ($k$) and fast heat removal from the reactor core, if necessary. A crucial technical factor is a long thermal response time which will make runaway processes such as occurred at TMI and Chernobyl vanced "Gen. III+" designs such as the Modular Pebble Bed Reactor (MPBR). India has commissioned a nuclear reactor utilizing thorium (Th), rather than uranium, as primary nuclear fuel, a strategy that could further defuse several of the Strategic Issues associated with nuclear energy. The country is building an entire new nuclear power industry upon this principle.

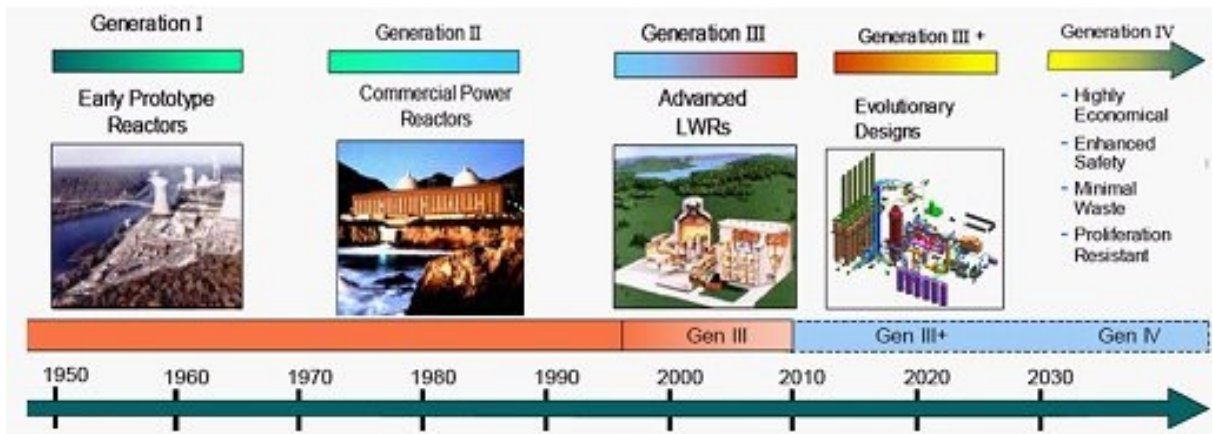

Figure 14: Timeline for evolving reactor generations Gen I – Gen IV characterized by improved operational and proliferation safety at reduced waste. In some countries, Gen III+ reactors are already under construction, ahead of this schedule (Adapted from DOE-Gif.)

Within the GNEP framework, the U.S. considers longer-term (next 20 years) several, economical Gen. IV designs (DOE-GIF, 2002), to be essentially impossible. Because of the hermetic enclosure of the reactor vessel within several massive containment shells and a separation of primary from secondary cooling loops, all new reactor types promise effective capture of all radioactive waste, including gases (*Kr, Xe, etc.*) produced in fission.

The fast (fast-spectrum) reactors in the above list are "breeders." They are all based on fission of *natural* uranium and of the successor materials bred in fast neutron reactions. Here, both isotopes, $^{235}$U and $^{238}$U, are reacted with energetic neutrons, releasing its entire nuclear energy potential. Thorium/uranium mix is an interesting alternative breeder fuel (David, 2007). Since the process is induced by fast neutrons, it does not require a moderator. The "top-ranked" (DOE-GIF, 2002) LFR can be configured as an assembly of prefabricated reactor modules rated at 300-400 MW$_e$. The small individual cores have a long lifetime of 10-30 years. The LFR safety is enhanced through its use of a very inert, lead or a lead/bismuth (Pb/Bi) eutectic coolant, which is liquid under normal operational conditions. The LFR has a closed fuel cycle and allows for efficient actinide management (see section below). This project is not yet pursued in the U.S. where the development of a helium-cooled VHTR has higher priority.

The VHTR project connects to a successful series of experimental reactor tests carried out between 1966 and 1988 in Germany (AVR and KFZ Jülich Pebble Bed Reactors) and the current modular developments in Pretoria/South Africa (PBMR). A "Pebble Bed" reactor is operated at core temperatures near 1000$^0$C, or above, and at pressures of up to 9 MPa. The uranium fuel consists of uranium oxide (UO$_2$) particles coated with silicon carbide enclosed in tennis-ball sized graphite spheres. A typical load would contain 450,000 of these pebbles. The reactor core con-cept has been tested successfully in the 21-year operations of the AVR reactor.

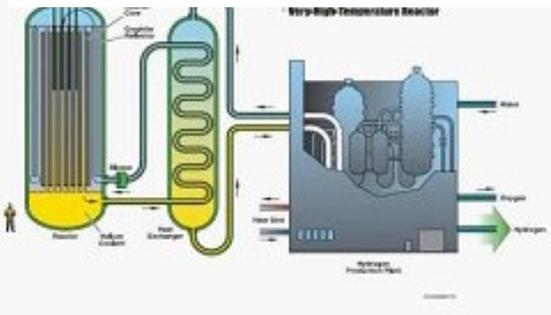

Even though operating temperatures are high in a PBR, its power density is 30 times lower than in a conventional reactor. The PBR demonstrates an example of an intrinsically fail-safe system that cannot be circumvented by human intervention: The ratio of surface area to volume of the fuel elements (pebbles) in the core is so large that heat is emitted through the surface much faster than can be produced by the fission fragments in the fuel. Therefore, heat build-up damaging the fuel elements ("melt-down") cannot occur, regardless of the presence or state of the coolant. The reactor has a strongly negative coefficient of reactivity.

Major goals of the VHTR project (see Fig. 15) are to develop by 2020 a reactor able to produce simultaneously both electricity and hydrogen by thermo-chemical dissociation. How important the latter capacity will be remains to be seen. The economic prospects of a "hydrogen economy" have been critically discussed by Romm (Romm, 2005) and Smil (Smil, 2005). In the context of reactor safety, it is mentioned here that the new class of subcritical, accelerator driven systems, which are currently not pursued in the U.S. are inherently incapable of meltdown. Furthermore, "breeder reactors" producing nuclear fuels from fertile isotopes are in their safety features identical to those of the underlying functional scheme of the host design.

The sodium (Na)-cooled fast reactor (SFR) is a system developed in the U.S. as a prime candidate for a "burner" or "incinerator" reactor for the high-level waste management. This type, which uses a closed fuel cycle, has already been built and operated safely in the U.S., Japan, France, Germany, Russia and other countries. The reactor derives its specific safety feature from the fact that the un-pressurized metal coolant has a large thermal inertia. Its boiling point is 883$^0$C, and the core typically runs at about 500$^0$C. There are two sequential Na cooling cycles, such that a potential reaction of Na with external water does not involve radioactive Na. The reactor can run on uranium-plutonium mixed-oxide UPuO$_2$ (MOX) fuel, on the waste from conventional reactors, or on a U/Pu-zirconium alloy, which would be difficult to convert into weapons material. In these and all reactor types considered by the GNEP initiative, reactor safety plays a most important role.

### B.    Resource limits of nuclear fuel

Like fossil fuels, nuclear fuel is a finite resource. About 10,000 t of uranium is used per year to produce electricity. Uranium is found mainly in Australia, Kazakhstan, Canada (Cigar Lake), the U.S., also in Africa (South Africa, Niger, Gabon, Namibia). In the U.S. uranium mines are in Wyoming, Utah, Colorado, New Mexico, and Texas. Presently, economically recoverable uranium reserves amount to approximately 4.7 Mt (IAEA, 2005). This supply can satisfy demand by conventional, wastefully inefficient, nuclear reactors for 50-100 years (Herbst & Hopley, 2007, Nifenecker, 2003), while new "smarter" breeder reactors would run on the same fuel for more than a thousand years. Following the present scenario of rising uranium prices, more uranium resources will be exploited. This already includes the reopening of mines and the mining of ore with poor uranium concentration. In future it could include seabed mining and uranium recovery from coal.

Uranium is extracted chemically from the mined uranium ore and converted, first to $U_3O_8$ oxide ("yellow cake") and then to uranium hexa-fluoride ($UF_6$). The latter is used in gas diffusers or centrifuges to enrich the gas stream in $^{235}UF_6$, which finally reaches 3-5% enrichment in its $^{235}U$ component. An important alternative is the mixed-oxide MOX fuel produced by blending or during a recycling process.

At present, mined uranium is needed in the U.S only to provide 55% of the needed primary reactor fuel. The rest comes from decommissioned nuclear weapons stockpiles, which is highly enriched material that can be blended in with natural uranium or thorium. Approximately 174 t of weapons-grade uranium have been made available for the civilian sector, 30 t have been used already, about 10% of the global demand (Herbst & Hopley, 2007). Fuel from this resource will continue to be available for some time. This fact has an influence on the economy of nuclear energy in the U.S. Thorium $^{232}Th$ is an interesting alternative fuel, which is also more abundant than uranium.

For a sustainable, long-term development of nuclear power in the U.S, it is necessary to abolish the current, wasteful "once-through" practice of using only the rare uranium isotope

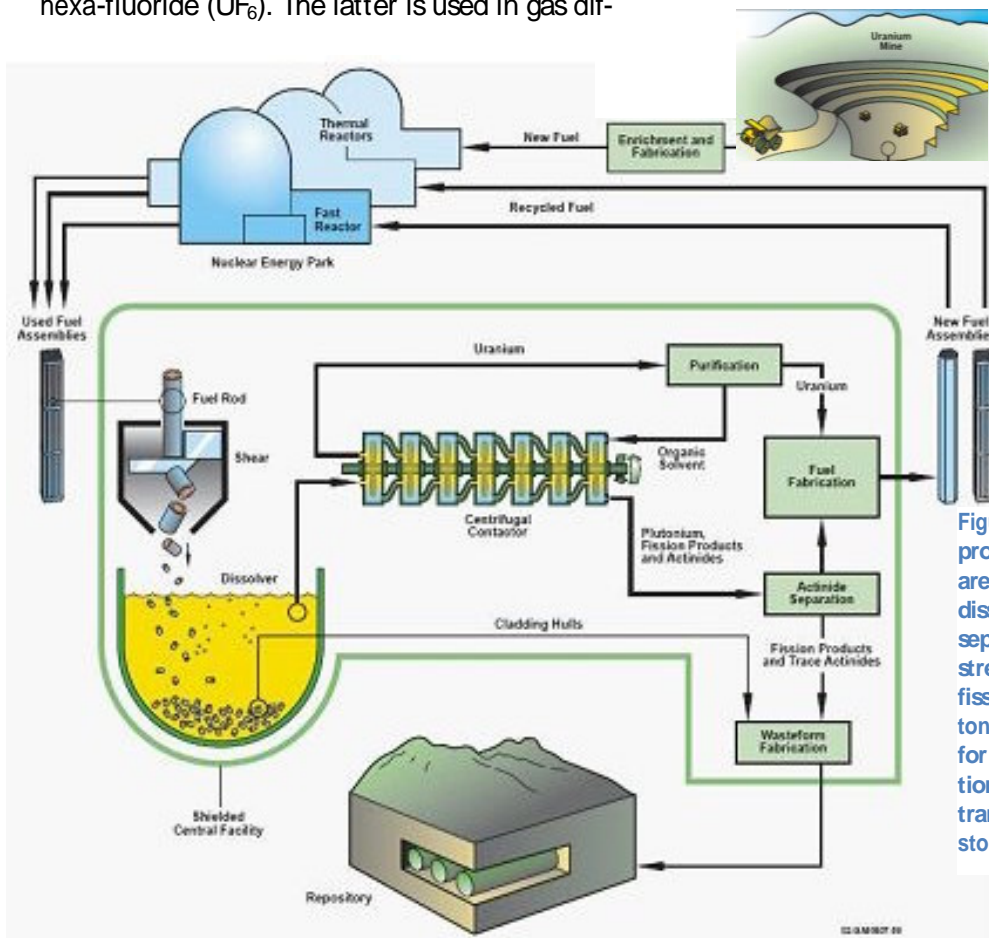

Figure 16: Aqueous nuclear fuel reprocessing scheme. Spent fuel rods are cut open and the contents are dissolved chemically. An isotopic separator provides different streams for uranium, actinides and fission products. Uranium and plutonium are recast into fuel elements for another cycle of reactor operation. Fission products and some transactinides are vitrified and stored in a long-term depository.

$^{235}$U (0.7% abundance) as fuel, and only a fraction at that. The once-through method produces relatively significant amounts of high-level radioactive waste while leaving 99.3% of the energy content of the natural uranium untapped. Instead, in the GNEP a closed fuel cycle such as illustrated in Fig. 16 will be adopted, where the "spent" fuel is chemically reprocessed, recovering in the so-called UREX, UREX+, or PUREX processes the uranium or the combined uranium/plutonium isotopes that remained after a long run or were "bred" in the preceding cycle. In these repeatable processes, fission products, which can act as reactor "poison" and perhaps some long-lived transuranium isotopes are removed from the reusable fuel elements. If fuel is recycled for use in a fast ("Burner") reactor, transuranium isotopes are left in the fuel to undergo transmutation inside the reactor while producing energy.

Depending on the new (Gen IV) reactor type considered, different reprocessing methods are employed. For example, an aqueous process is preferred for the spent fuel from an SFR or an LWR working with MOX fuel, while pyroprocessing is the method of choice of that reactor when loaded with metallic fuel (DOE-GIF,

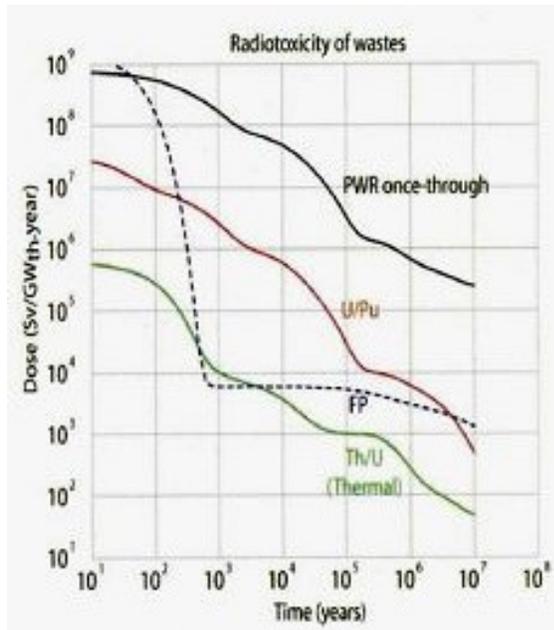

**Figure 17: Radio toxicity vs. time after shutdown, of spent fuel from a pressurized water uranium reactor (PWR), from a U/Pu breeder, and from the Th/U fuel cycle. FP indicates the faster decay of fission products. From (David, 2007)**

2002). The former type of reprocessing/recycling process is well known from many years of experience in other countries. However, to enhance the economy and reduce the ecological impact of this reprocessing method, further R&D is desirable. The pyro process has been employed in the past, but several steps in an advanced version are still under active development.

Many countries such as the U.K., France, Russia, and India have applied nuclear fuel reprocessing in a closed fuel cycle, safely and for a long time (the U.K. since the 1950s). France and Russia provide the service commercially to other countries such as Japan and Germany. Present annual capacity of reprocessing plants is approximately 5,000 tons. An environmental concern relates to undesirable consequences of fuel reprocessing, accumulating high-level radioactive waste that has to be managed. In the 1970s, the U.S. adopted a non-reprocessing policy and closed down its reprocessing plants. The intended change to a closed-fuel cycle nuclear industry now requires technology import from European countries and/or India, specifically for reconstituting lost expertise in nuclear and radiation chemistry.

Obviously, technologically well understood fuel breeding, reprocessing and recycling extend the time span of available nuclear energy for *thousands of years* at present energy demand levels. This estimate does not even account for the thorium (Th) fuel which is (~4x) more abundant in the earth's crust than uranium and would in *several* respects represent a more desirable fuel technology. In any case, the ample fuel resources available already with present technologies make nuclear energy the prime candidate for a major component of a long-term energy strategy. Strategic Issue B has been resolved.

### C.  Radiotoxic waste management

A major argument leveled often against the use of nuclear power is the fact that the operation of conventional nuclear power plants produces copious amounts of high- and low-level radiotoxic materials as byproducts of the fission/breeding process. These trace materials contaminate (~1% of total) the "spent" $^{235}U/^{238}U$ fuel elements that are presently stored on site at most nuclear plants.

The radioactivity of fuel rods freshly removed from a core is essentially due to fission products (FP). Due to their short half lives, this activity disappears within 6 months (Stacey, 2001).During the first three years after removal from the reactor core, spent fuel elements are intensively radio-toxic (Fig. 17), such that they are kept in water pools to "cool." After that, passive cooling in dry storage is appropriate (APS, 2007). The corresponding volume is not large (2 $m^3$ per 1-GW reactor each year). An amount of 54,000 tons of spent fuel rods has accumulated so far on the world's reactor sites. The "waste" would physically fit into one corner of a large U.S. commercial warehouse for consumer goods, although such a concentration would be inadvisable.

The fission byproducts of nuclear power generation release energy in form of alpha, beta and gamma rays for extended periods of time. The former particles are examples of directly ionizing radiation of the type acting in nuclear batteries. Until large-scale, economic applications for these secondary energy sources are found, the materials are considered undesirable and dangerous. Funds have been collected by the U.S. federal government from mandatory fees of 0.1 cent imposed on every kWh of nuclear generated electricity, contribut-ing approximately US$ 1 B/year to a fund dedicated to the development of a geological nuclear waste repository in the country.

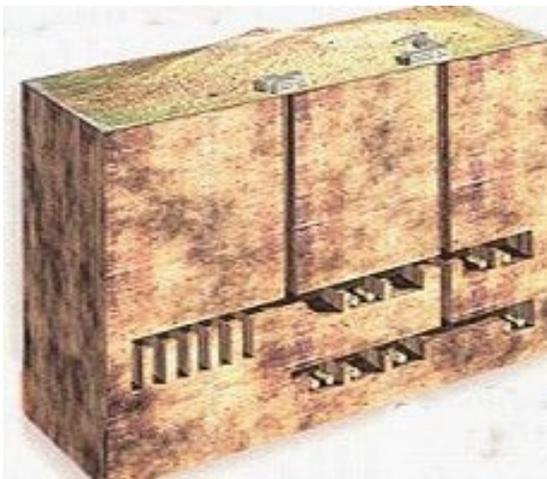

Figure 18: Artist's view of underground waste depository showing service buildings and shafts, and storage caverns.

In the terminology of the energy sector, unlike the waste of fossil fuel power plants, nuclear waste has always been captured and mostly stored on site. But the material has not yet been sequestered in a permanent geological repository such as in salt domes or geologically stable rock formations. In the U.S. this latter possibility has been discussed and developed over a span of many years, until very recently (2007) the Yucca Mountain site in New Mexico was designated the first official permanent nuclear waste repository, to open in 2010. The site choice and details of environmental engineering study and selection process have become contentious political issues. In addition, concerns for the safety of transporting nuclear waste are often raised, even though the safety of the thousands of past shipments moving 3,500 tons of waste through the U.S. has been documented (NAS, 2006).

It has been argued earlier in this article that, in fact, there is no convincing evidence for the soundness of predictions of a geological safety of any such repository over periods of millennia or even centuries. In addition, what qualifies as waste now may (likely) have value in future. Therefore, it appears to be prudent to design repositories such that they allow for a maximum of flexibility in dealing with the stored waste at any time in the future, including extraction and relocation, recovery and repackaging.

The basic waste removal problem faced by the nuclear industry applies also to the fossil fuel energy sector, which has mounted carbon sequestration efforts. However, the corresponding problems for nuclear energy are different and somewhat less severe:

- The radio-toxicity of nuclear waste disappears in time.
- There is no known positive feedback effect that could generate a runaway situation similar to what has been postulated for climatic influences of GHG emissions.
- Radioactive waste is immobile, since it is mostly in solid (vitrified) and less in liquid

form, while gas components rare and mostly short-lived.

- Any long-term discharge from a repository would affect area and population of limited sizes.

Lastly, a new policy of reprocessing nuclear fuel reduces considerably the demand for storage space, decreasing the required residence time by several orders of magnitude (see Fig. 17).

The main tasks in nuclear waste management is to separate radio-chemically the reusable U or U/Pu fuel stock from fission products (FP) and all, or some, of the transactinides produced in multiple neutron captures. To effectively transmute these products, for example by neutron irradiation, requires prior element separation. Target for removal are the relatively long-lived fission products $^{99}$Tc ($t_{1/2}$ ~$2 \cdot 10^5$ years) and $^{129}$I ($t_{1/2}$ ~$1.6 \cdot 10^7$ years). The gamma emitting FPs $^{90}$Sr and $^{137}$Cs decay both with $t_{1/2}$ ~$30$ years. The long-term activity of spent fuel is dominated by $^{240}$Pu and the "minor actinides" (MA) Am, Np, and Cm. While effective commercial separation techniques are available for Pu, such methods have still to be devised by radiochemists (Stacey, 2001). As illustrated in Fig. 17, the effect of these actinides can be reduced significantly, down to 0.5% of the initial activity, by their recycling as fuel in fast ("Burner") reactors of the SFR type planned in the GNEP initiative.

Invoking the Th/U breeder fuel cycle would essentially defuse the nuclear waste problem. As demonstrated by the radio-toxicity data shown in Fig. 17 (David, 2007), radioactivity levels of reprocessed fuel due to transactinides from the Th/U cycle reduce within less than one hundred years to a level approached by waste from conventional (PWR) reactors only after a million years. Its radio-toxicity decays within similarly short time scales similar to those of fission products (FP) as illustrated in Fig. 17. This is clearly an attractive feature of the Th/U fuel cycle employed by the Indian nuclear energy program (Dey & Bansal, 2006; Sinha & Kakadkar, 2006). It should justify similar considerations of this fuel cycle elsewhere.

In summary, the need for long-term geological deposition of some nuclear waste has not disappeared and will remain for all considered nuclear fuel cycles. However, its severity is significantly smaller than typically assumed by the public and compares favorably with the environmental hazards accepted by other segments of the energy economy. Strategic Issue C has thus lost much of its severity and urgency.

### D.   Proliferation safety

The potential for creating powerful nuclear weapons with relatively small amounts (~10 kg) of concentrated (95%) uranium (isotope $^{235}$U) or plutonium (isotope $^{239}$Pu) adds a risk factor unique to the nuclear industry. Even a small reactor produces sufficient amounts of plutonium to manufacture several fission ("atomic") bombs per year, albeit typically in a highly diluted isotope mix. However, to optimize the production of materials suitable for weaponization, a reactor would have to be shut down, discharged, refuelled, and restarted several times per year in a way quite obvious to external observers. Nevertheless, isotopically separated and recovered from reactor fuel, some of the $^{235}$U and $^{239}$Pu could be captured and diverted by non-government agents who then could produce a simple "home-made" nuclear device.

Reprocessing spent nuclear fuels with current reprocessing technologies requires the transfer of large amounts of Pu or enriched materials from reprocessing facilities to reactors. The risk of illicit interceptions of such materials in transit has conjured the spectre of a perilous "plutonium economy." However, this risk could easily be reduced by collocating reprocessing laboratories with the nuclear reactor facilities using the fuel. Given the potential of social unrest, a proposed alternative strategy of placing reprocessing plants only in those countries that are already fuel suppliers, is less convincing.

A proliferation-safe strategy adopts a fuel cycle that simply does not produce highly enriched nuclear fuels. In fact, several new reactor types mentioned earlier are designed for fuel types and reprocessing/recycling techniques that are "proliferation-safe" in that they make isolation and diversion of fissile materials extremely difficult, requiring access to a sophisticated metallurgical national industry and infrastructure. This result can be achieved in various ways that are under active consideration. For example, alloys instead of pure metals could be used as fuels. Already laboratory tested, the Uranium Extraction Plus (UREX+) reprocessing technique achieves these goals by keeping uranium, plutonium and the minor actinides (Np, Am, Cm) together in the process. In this fashion, fuel enrichment levels are too low for weapons' applications, greatly reducing the proliferation risks. As a side benefit, the actinides would be taken out of the waste stream and provide new fuel, for example, for a "burner" reactor.

It should also be recalled in this context that issues related to proliferation are much less severe for the Th/U fuel cycle and accelerator driven subcritical (ADS) reactors. Isolating weapons grade fissile materials from spent Th/U fuel elements already requires sophisticated remotely controlled radiochemistry, which is made unfeasible by natural uranium admixtures to the basic Th fuel. Both these potential nuclear techniques are under study and development in several countries.

GNEP decisions about recommended future reactor types and the fuels they use are weighed heavily by assessments of the relative advantages of fuel reprocessing against the proliferation risks. Because of the large number of countries that presently already conduct, or intend to pursue, fuel reprocessing, combating nuclear proliferation remains a recognized international safeguarding task with no patent solutions.

The proliferation risk is now understood to include the possibilities of "dirty" bombs, which could be designed to spread radioactive materials over populated areas using conventional explosives. It should be noted that the materials that have been in storage for some time at reactor sites are not the preferred substances for such purpose. Because of the long half lives of the isotopes in spent fuel rods, the material is less radiotoxic. It would also be more readily cleaned up after an incident.

It is important to distinguish the narrow concept of "proliferation" adopted here from its official political definition, which includes the transfer and sharing of knowledge about nuclear energy technologies between legitimate governments of nation states. The narrower definition recognizes that the denial of advanced technologies to entire nations by monopoly holding powers may be ethically untenable. History demonstrates that such attempts are also futile. Present international treaties notwithstanding, a number of nations, including Israel, China, India, Pakistan, North Korea, and Iran have gained access to nuclear energy technologies and have acquired the capability to produce nuclear weapons, thus enlarging the circle of nuclear powers significantly. Yet, in spite of much international tension and a number of armed conflicts, in 62 years, the U.S. has remained the only nation that has ever used nuclear weapons. Simply, a policy designed to maintain nuclear exclusivity among nation states has failed, but without disastrous consequences...yet. A realistic view of international relations that acknowledges that fact would decriminalize, and even promote, a transparent exchange of information pertaining to nuclear power between legitimate governments.

Summarizing, one arrives at the conclusion that the nuclear proliferation risk, narrowly defined as most people would understand and accept, has become very small, even with existing reprocessing technologies and with minimal international surveillance (e.g., by the IAEA) of reactor operations. It promises to become essentially eliminated by new technologies now explored in laboratory studies. Together with answers to the questions discussed earlier, this conclusion also implies that an important Strategic Issue associated with nuclear energy could be considered nearly resolved, as far as technologically possible. What remains can be resolved by the political process. This is an important fact that the public should be made aware of.

## E.  Economy of nuclear energy

In previous sections, costs of electricity generated from different technologies have been discussed. For most applications, including nuclear power plants, comprehensive life cycle cost/benefit analyses are unavailable. In addition, these costs are time dependent. Capital and fuel, as well as collateral costs, are quantities that fluctuate strongly with supply and demand, as well as with the exchange rates of the currencies involved. However, the costs of nuclear and coal generated electricity are comparable to each other (4-5 cents/kWh) and amount to about 10% of the costs of renewables. In both cases, the "energy payback" time (energy produced=energy spent in plant construction) is only a few months, compared to several years for renewables. Plant capital and operating costs are also comparable (between 3 and 5 cents/kWh).

Costs for final sequestration/deposition of the waste are not well known, for either fossil or nuclear fuels. Some fee for waste disposal has already been applied to nuclear electricity. With the adoption of closed fuel cycles, the projected amount of nuclear waste that will eventually have to be placed in a permanent depository is reduced. New fuel is generated by reprocessing or "breeding." This fact improves the cost situation further in favor of nuclear power. Furthermore, in the U.S. the availability of weapons-grade uranium for civilian purposes reduces fuel costs at least during a decade.

The costs of complete decommissioning of nuclear plants would certainly be expected to be higher than that of coal power plants and more in line with that of large chemical, refinery or hydro-electric facilities. However, in the present and ongoing expansion phase, collocation of new nuclear reactors with existing sites will allow for a reuse of much of the technical infrastructure and plant already in place.

Economic considerations are important for the design of a sustainable energy strategy. But it has become common sentiment that they should not be the only or even the dominant ones. The specific additional collateral costs in human health and lives of different technologies are much more important but difficult to quantify. However, because of the absence of significant emission of airborne pollutants, toxic liquids, and GHGs, nuclear power is already now "cleaner" than other energy technologies, besides being economically competitive. This is no longer a Strategic Issue.

### F.    The future of nuclear energy research

As discussed above, the economy can rely on a set of already proven and tested nuclear energy technologies to support its ongoing expansion. This poses the question of whether nuclear science is a field that, after almost 70 years, has already grown to maturity with little significant potential left for scientific discovery that could be of relevance to energy applications.

The answer to this query has to be that nuclear science is evolving and expanding. It remains relevant to a variety of applications, including energy. Projects of basic science interest deal with fundamental processes of energy conversion, which should have a positive feedback on energy technology.

Nuclear science has developed an understanding of basic properties of nuclei in the neighborhood of naturally occurring nuclei ("Valley of Beta-Stability") and for low excitations. But it is struggling to gain theoretical access to phenomena of the stability of nuclei with unusual combinations of neutrons and protons ("isospin") (Li & Schröder, 2001) or of nuclei with high internal excitations. Their large-amplitude fission-like motion and their interactions remain topics of current research (Hofmann, 2007).

Even though it has been possible to parameterize and systematize large sets of nuclear data within a conventional theoretical framework, the difficulty to extrapolate to properties and interactions of weakly bound, exotic nuclei suggests that some major pieces of the physics puzzle are still missing. To understand this physics well and quantitatively belongs to the prerequisites for the success of basic energy science in the future. It is of considerable interest for obtaining a better understanding of cosmologi-cal structures (neutron stars) and processes

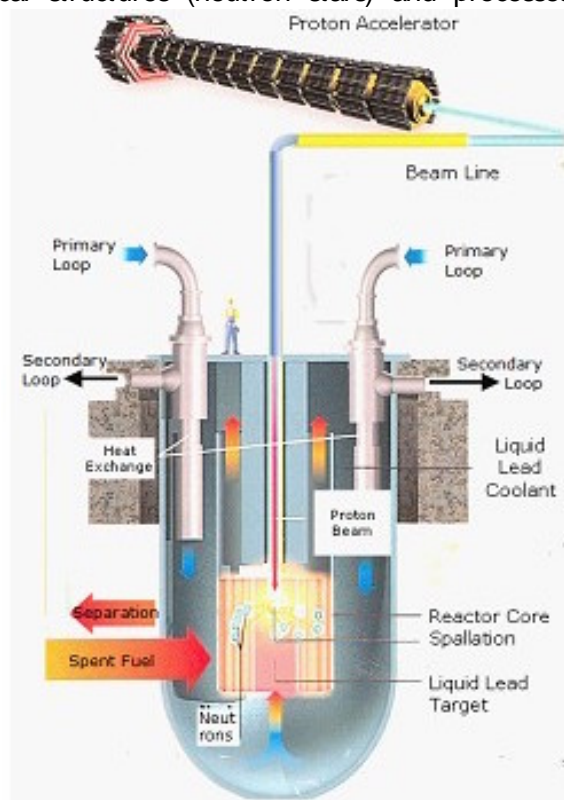

Figure 19: Subcritical accelerator driven transmu-
ter/burner reactor (Nuclear Diesel). The proton beam
produces fast neutrons in a liquid lead target, which in-
duce transmutation and fission reactions in the reactor
core proper.

(supernovae) as well.  In fact, the fundamental
role that the nucleus and nuclear phenomena
play for an understanding of processes on mi-
croscopic or cosmological scales provides strong
motivation for much of modern nuclear re-
search.

As an example of the importance of basic nuc-
lear research for future energy applications, one
may consider (Fig. 19) the accelerator driven
subcritical (ADS) nuclear reactor (Bowman
1992, Rubbia, 1993, Andriamonge, 1995, Nife-
necker, 2003) which could also be used for
transmutation of transactinides and fission
products. In the basic process, a beam of high-
energy (1 GeV) projectiles, protons or neutrons,
produced with an external accelerator is used
to bombard the "target" material in the reactor
core. The projectiles induce a chain of multiple
successive (primary, secondary, ternary, etc.)

spallation, fission, and evaporation reactions. As
may be deduced from Fig. 12, the overall
process is exothermic. This fact makes it inter-
esting for nuclear reactors that can be fuelled
with a range of heavy materials including, but
not limited to, mercury, lead, bismuth, thorium,
uranium ("nuclear Diesel" reactor), where the
lighter materials serve mainly as neutron gene-
rating "additives" to the fissile fuel base.

Perhaps surprisingly, already the first two
steps in the reaction sequence, the formation of
a heavy target-like nucleus (target nucleus + 1
projectile nucleon) and its decay pose formida-
ble, fundamental challenges to conventional
nuclear reaction theory (Schröder & Huizenga,
1984). The reaction reaches excitation energies
of a few MeV per nucleon (mass unit), well be-
low the binding energy of approximately 8 MeV
per nucleon (Fig. 12). The inability of theory to
account for experimental observations of unex-
pected branching between evaporation, fission,
and copious cluster emission in heavy-ion reac-
tions has opened a new direction of nuclear
science research, the physics of nuclear phase
transitions. The anomalies with respect to expe-
rimental systematics of multi-chance fission es-
tablished (Hilscher & Rossner, 1992) for lower
excitations occur consistently (Töke & Schröder,
2006) in both heavy-ion (Schröder, 2003) and
spallation (Goldenbaum, 2004, Tishchenko,
2005) reactions. A satisfactory explanation for
these processes will probably require a much
better understanding of the diffuse surface
(Töke & Schröder, 2003) of excited nuclei than
has been provided by contemporary nuclear
physics. Hence the motivation and need for
studies of the nuclear surface about which sur-
prisingly little is known.

To quote another example for a basic re-
search need, nuclear and radiation chemistry is
challenged to provide a better understanding of
the chemical processes that will make the mol-
ten-salt reactor (MSR) work, which is one of the
elegant advanced nuclear reactor types fore-
seen by GNEP in the medium-term future. This
system (see Fig. 20) uses a circulating molten
salt mixture as fuel and coolant. For example,
uranium, thorium, and actinide fluorides are dissolved in a molten mix of sodium and zirconium fluorides (NaF/ZrF$_4$).

Whether or not this would actually be a preferred solution is subject to favorable chemical kinetics and thermodynamics of the molten salt mix. Solubility of actinides and fission fragments in the salt mix needs to be measured. Boundary conditions are given by the radiochemical requirements of low neutron capture cross sections of the passive components and their transmutation in an intense neutron flux.

In addition to basic nuclear science research of the type illustrated with the above examples, there is great need for new and more precise and detailed measurements of nuclear cross section data (Schroeder & Lusk, 2006). Correspondingly, a more accurate nuclear reaction theory and large-scale computer simulations are needed. Such research is of importance for reactor systems such as fast-spectrum reactors considered in the GNEP project (DOE-GIF, 2002, Nifenecker, 2003, Aliberti, 2006). With secondary beams of exotic projectiles that have become available, or are under development, in Japan, Europe, and in the U.S., one can now study interactions of the kind that fission or spallation products undergo in a reactor core.

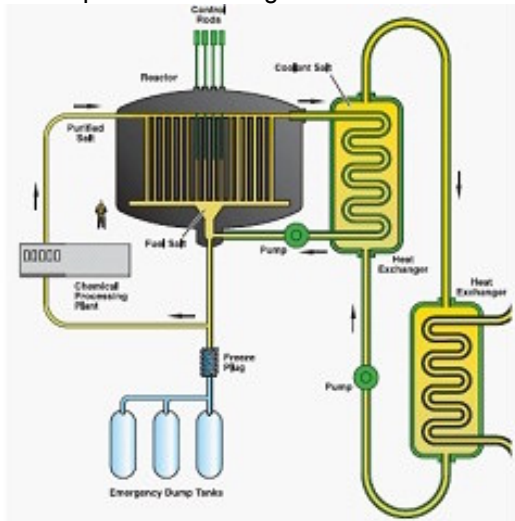

The importance of new R&D in support of the GNEP initiative has been recognized by the corresponding government agencies. Already since 1999 a U.S. Nuclear Energy Research Initi-ative (NERI) has supported nuclear engineering projects for Gen IV and the Advanced Fuel Cycle (AFC). NERI research partners in 2006 included 36 U.S. university nuclear engineering departments, eight national laboratories, along with several foreign organizations and private corporations. Founded in 2001, an international cooperate (I-NERI) supports joint nuclear energy research in collaboration with Asian, European, South American and OECD countries.

Since more recently, the DOE sponsored Advanced Fuel Cycle Initiative (AFCI) encourages the formation of collaborations between basic (physics and chemistry) and applied nuclear scientists and engineers, to find strategies for generating data and theoretical models of relevance to reactor of Gen IV and beyond. Supporting AFCI objectives are further several computer and engineering laboratories, all planned to take up operations within the next few years.

In the discussions of the goals and aspirations of GNEP and other initiatives interested in nuclear technology, it has been recognized that over the past several decades there has been a erosion in the number of nuclear science faculty and graduates at U.S universities. This has led to a loss of global competitiveness in a fundamental science and technology which is of strategic importance for the future strength of the U.S. economy.

## Conclusions: Recommendations for a sustainable energy future

The discussion of the previous sections has reiterated and commented on what is commonly known, namely that cheap oil and gas, the energy basis of the economies of Western industrialized nations, are rapidly disappearing. It is uncertain exactly how many decades of their affordable use remain. Certain is, however, that in the interim a replacement energy basis has to be found and adapted to an optimum extent to the requirements of modern economies. It is also obvious that, in turn, economies have to adapt to the specifics of the available energy re-sources. Reduction of demand, albeit a "smart" way to mitigate the effect of an energy crisis, is not a complete and certainly not a global solution to, or a way to circumvent, the problem. A number of populous emerging nations will simply need to increase their energy consumption, largely and necessarily based on fossil fuel technologies. On the other hand, major and sudden reductions in living standard would be difficult to establish in other nations. But the technologically advanced societies could be expected to realign their economies with nuclear and clean-coal electricity.

The remaining time afforded society by the currently still relatively low energy prices should be used to evaluate critically and realistically the available technological choices for energy futures, and then to develop vigorously the most potent technologies. The large numbers of government and privately sponsored research projects, as well as advertising efforts of a burgeoning new industry, assert that there are many attractive energy futures for the industrialized nations. In fact, a comparative evaluation of potential and risks of several primary energy technologies, such as attempted in this article, suggests otherwise. Several energy futures that at first appear attractive ("green") and painless to implement do not actually live up to their promises or fulfill essential requirements that economies make on any sustainable energy basis.

**Reliable primary energy generating technologies are based on the utilization of clean-coal and advanced nuclear energy**, neither of them first choices of the ecologically concerned. However, with these two technologies, existing economies can evolve smoothly by gradual extrapolation and improvement of their existing infrastructures. The main criticisms against these technologies, as represented by the corresponding *Strategic Issues*, have already been, or can be, addressed with existing methods. Required are public support and commitment to invest heavily in the future energy infrastructure. At the same time, it is important to continue the R&D of renewable and fusion energy technologies, which could play important roles in the longer term.

Any realistic strategic energy plan for a (relatively) secure energy future would consider, among others, with priority the following set of specific RD&T recommendations:

**Develop and Employ Clean Coal Technology:**
- Improve the safety of mining (robotics).
- Enforce environmental restoration of mining sites.
- Promote and subsidize clean coal (IGCC) technology. Eliminate airborne sulfuric, GHG, PM, and metal emissions.
- Further develop, test, and construct modern FutureGen-type coal power plants, upgrade existing fossil-fuel plants to multi-fuel capability, including coal.
- Develop a new and efficient chemistry of synfuel conversion of coal.
- Reassign coal power plants to produce synfuels for transportation.
- Conduct large-scale/long-term CCS tests to find semi-permanent, monitored $CO_2$ repositories.

**Develop and Employ Advanced Nuclear Power:**
- Continue to improve the intrinsic safety of nuclear reactors and processing plants.
- Test and construct advanced modular nuclear reactors near sites of existing plants.
- Test and construct advanced burner/transmuter reactors to reduce radiotoxic waste.
- Import and develop further closed nuclear fuel cycle technologies.
- Develop and test proliferation-safe reprocessing methods (e.g., UREX+).
- Test and develop a closed Th/U breeder fuel cycle.
- Develop ADS systems and the required high current accelerator technology.
- Develop the chemistry of molten salt mixtures, and design a molten salt test reactor.

- Expand the radio-chemistry of actinides, transactinides and fission products.
- Start operating a semi-permanent nuclear waste depository with a flexible strategy.

Obviously, to achieve the goals of these two sets of recommendations, one needs to emphasize much-discussed traits of a modern "information society" with specific application to energy. However, this intellectual development would remain insufficient without revitalizing domestic industrial infrastructure, where it has been lost. To conduct such an energy upgrade program successfully, higher demands have to be made on R&D in coal and synfuel chemistry, in chemical engineering, in nuclear and radiation-chemistry, nuclear engineering, as well as in low-energy basic nuclear science. To restore expertise lost in past decades, stronger efforts have to be made to increase the number of university faculty and graduates of nuclear science much above the current low levels.

University communities bare a large responsibility to analyze strategic energy problems facing industrialized society and to engage in working realistic solutions. They should be made aware of the associated opportunities in basic and applied research in nuclear science and clean coal technology as meaningful and interesting professional careers. Government agencies have corresponding responsibilities:

**Educate and Train Personnel**

- Promote and fund with priority energy science and technology development.
- Recruit university students and faculty to embark on careers in basic and applied energy science and engineering, emphasize nuclear science/technology and chemical engineering.

Putting these issues into a larger reference frame, one realizes the true scale of the tasks. The eventual implementation and realization of the above recommendations require a strong and competent workforce running an upgraded domestic manufacturing ("brick-and-mortar") industry and an efficient, modern infrastruc-

ture. A sustainable energy strategy cannot be realized, unless the following tasks are also accomplished:

**Upgrade Industry and Infrastructure:**

- Invest in and promote domestic machining and manufacturing industry, specifically the capacity to engineer, construct and operate advanced modular power plants, generators and distribution networks.
- Promote new and electrified public transport/rail systems.
- Develop new electric grid and other distribution technologies.
- Develop new strategies for grid expansion and install a new nationwide grid.
- Develop new energy storage techniques.
- Stratify the power plant regulation process without compromising public safety.

The inevitably required massive public investments (of the order of a trillion $ per decade) in an improved energy infrastructure and the associated gradual transformation of the economy can be pursued long term only with the consent and full cooperation of an informed public. Therefore, it is necessary to invest in public education regarding energy issues.

**Inform the Domestic and International Public:**

- Inform the public about realistic potential and risk of energy technologies.
- Promote reduction in energy demand, increase energy efficiency.
- Invest in and promote science education at all levels of education.
- Devise a new international compact about the use of advanced vs. traditional energy technologies by countries with compatible technological infrastructure.
- Devise a new international compact on the sharing and control of knowledge about nuclear technology, avoid monopolies of knowledge, and decriminalize knowledge sharing between nation states.

Undoubtedly, the coming decades will witness a large-scale reformation and realignment of Western industrialized societies. Guided by perceptive, courageous leaders and supported by an informed and concerned public, a gentle transformation could be achieved with multiple concurrent programs, each of a magnitude dwarfing the U.S. Manhattan Project. Taking on this historic task with courage and determination would be most effective now, when there is still time. The venture would also be a source of inspiration for generations to come.

### Acknowledgments

The author has appreciated comments and questions by Drs. M. Blann, D. Douglass, and D. Hilscher on the first draft of this article.